# Dimensionality Enhanced Out-of-Plane Spin Currents in NbIrTe$_4$ for Efficient Field-Free Switching of Perpendicular Magnetization

Wei Yang[1,2,3,†], Xinhe Wang[2,†], Jianing Liu[2,†], Daming Zhou[2], Xiaoyang Lin[1,2,*], Ke Zhang[2], Héloïse Damas[3], Xinyue Wang[4], Xianyang Lu[4], Haozhe Yang[2], Stéphane Mangin[3], Sébastien Petit-Watelot[3], Michel Hehn[3], Albert Fert[2,5], Juan-Carlos Rojas-Sánchez[3,*], Weisheng Zhao[1,2,*]

[1] National Key Laboratory of Spintronics, Hangzhou International Innovation Institute, Beihang University, Hangzhou 311115, China

[2] Fert Beijing Institute, School of Integrated Circuit Science and Engineering, Beihang University, Beijing 100191, China

[3] CNRS, Institute Jean Lamour, Université de Lorraine, Nancy F-54 000, France

[4] School of Integrated Circuits, Nanjing University, Suzhou 215163, China

[5] Laboratoire Albert Fert, CNRS, Thales, Université Paris-Saclay, Palaiseau 91767, France

[†] These authors contributed equally.

[*]Correspondence and requests for materials should be addressed to:

Xiaoyang Lin: XYLin@buaa.edu.cn

Juan-Carlos Rojas-Sánchez: juan-carlos.rojas-sanchez@univ-lorraine.fr

Weisheng Zhao: weisheng.zhao@buaa.edu.cn

# Abstract

Efficient generation of out-of-plane (OOP) spin currents is crucial for advanced spintronic memory applications. However, the theoretical understanding and experimental implementation of robust OOP spin currents for high-density and low-power magnetization switching remain significant challenges of spintronics. Here, we demonstrate that transitioning NbIrTe$_4$ from a two-dimensional quantum spin Hall insulator to a three-dimensional type-II Weyl semimetal markedly enhances OOP spin current generation. The bulk topological Weyl semimetal nature of NbIrTe$_4$, characterized by its Weyl cone, significantly enhances the OOP spin Berry curvature, enabling an unprecedented OOP spin Hall conductivity exceeding $10^5$ $\hbar/2e \cdot \Omega^{-1}m^{-1}$. This enhancement, surpassing the in-plane component by more than fourfold, enables efficient and field-free spin-orbit torque (SOT) switching of perpendicular magnetization with a low current density of 1.4 MA/cm$^2$. The improved spin Hall conductivity reduces the overall power consumption by more than two orders of magnitude compared to existing systems, such as heavy metals. Our findings highlight the pivotal role of dimensionality in harnessing robust OOP spin currents in topological Wely semimetals, paving the way for the development of high-density, low-power spintronic memory technologies.

# Introduction

Spin–orbit torque (SOT) has revolutionized spintronic device engineering by enabling electrical control of magnetization at reduced power consumption and without reliance on external magnetic fields[1-5]. Central to SOT-based applications is the generation of spin currents, which can be categorized into two main types depending on their polarization orientation: in-plane and out-of-plane (OOP). In conventional heavy metals with strong spin–orbit coupling, the spin Hall effect (SHE) produces a spin current whose flow, polarization, and driving electric field are mutually orthogonal [6-8]. For instance, an electric current along the *x*-direction generates a spin current polarized along *y*-direction ($\hat{\sigma}_y$), flowing along *z*-direction. While such an in-plane spin current $\hat{\sigma}_y$ exerts an SOT on a magnet with perpendicular magnetic anisotropy (PMA), an external magnetic field is typically required to break symmetry and achieve deterministic switching, hindering the scalability and practicality of SOT-based high-density memory applications[1,3,9].

OOP spin currents ($\hat{\sigma}_z$) provide a compelling alternative as the spin polarization aligns with the perpendicular magnetization, effectively counteracting magnetic damping and enabling efficient, field-free, and deterministic SOT switching[10,11]. The fundamental requirement for generating OOP spin currents $\hat{\sigma}_z$ is breaking lateral mirror symmetry[12-15]. As illustrated in **Fig. 1a**, when current is parallel to the lateral mirror symmetry plane, performing a mirror operation will invert the OOP spin current from $\hat{\sigma}_z$ to -$\hat{\sigma}_z$ without altering the current itself. This symmetry enforces $\hat{\sigma}_z = -\hat{\sigma}_z$ thereby nullifying the OOP spin current $\hat{\sigma}_z=0$. Consequently, orienting the current direction perpendicular to any existing mirror planes disrupts this symmetry constraint, allowing for a nonzero $\hat{\sigma}_z$ component, as demonstrated in WTe$_2$[10,16-19], TaIrTe$_4$[12,20], β-MoTe$_2$[21], Mn$_2$Au[22], MnPd$_3$[23], MnGaN[24], RuO$_2$[25] etc.

Despite the advantages of OOP spin currents $\hat{\sigma}_z$, the generation efficiency of $\hat{\sigma}_z$ remains substantially low[17,23,25-27], posing a significant challenge for the practical implementation of high-density spintronic devices. The strategies developed thus far primarily targeted interface and surface engineering of materials[13,14,28]. These approaches leverage materials with high spin conductivity to generate $\hat{\sigma}_y$-polarized spin currents, followed by low-symmetry materials to facilitate the transformation of $\hat{\sigma}_y$ to $\hat{\sigma}_z$ via mechanisms such as spin precession[14] or spin scattering[28,29]. However, the spin polarization inevitably diminishes during the spin-to-spin conversion process, significantly reducing the overall conversion efficiency. As a result, the spin Hall conductivity of OOP spin current ($\sigma_{sh}^z$) remain at least an order of magnitude smaller than the in-plane counterpart ($\sigma_{sh}^y$), underscoring the need for a more profound understanding and the development of robust mechanisms to generate and enhance OOP spin currents.

Here, we introduce a dimensional engineering strategy to enhance the OOP spin current generation in low-symmetry quantum material NbIrTe$_4$. By increasing the thickness of NbIrTe$_4$, the OOP spin Hall conductivity $\sigma_{sh}^z$ was enhanced fourfold due to enhanced OOP spin Berry curvature in bulk electronic states, reaching an impressive magnitude of $10^5$ $\hbar/2e \cdot \Omega^{-1} m^{-1}$. Notably, the OOP spin Hall conductivity $\sigma_{sh}^z$ surpassed its in-plane counterpart $\sigma_{sh}^y$, resulting in a substantial shift of the spin tilt angle from

~11° (in-plane dominant), to 77°, (OOP dominant). This remarkable enhancement in OOP spin current generation significantly reduced the threshold current required for deterministic magnetization switching in a PMA system, decreasing by an order of magnitude to ~1.4 MA cm$^{-2}$. Correspondingly, the power consumption was reduced by two orders of magnitude, providing profound insights for the development of next-generation energy-efficient spintronic technologies.

## Results and discussion

### Crystal Characterization of NbIrTe$_4$ Samples

The generation of OOP spin currents $\hat{\sigma}_z$ in NbIrTe$_4$ is closely related to its crystallographic and electronic structure properties. The T$_d$-phase of NbIrTe$_4$ crystallizes in the non-centrosymmetric space group *Pmn2$_1$*, forming a layered structure stacked along the c-axis, with a mirror plane perpendicular to the *a*-axis (**Fig. 1b**)[30,31]. As results, the generation of out-of-plane spin current is only permitted when the charge current flows along the a-axis of NbIrTe$_4$.

To investigate the spin current generated by NbIrTe$_4$, we deposited a Ti/ CoFeB / MgO /Ta stack as the ferromagnetic layer on NbIrTe$_4$ via magnetron sputtering (**Methods**). Following the deposition, the anisotropic crystallographic symmetry of NbIrTe$_4$ was confirmed through Raman spectroscopy. As shown in **Fig. 1c**, the Raman intensity at a shift of 151.67 cm$^{-1}$ exhibits a high integrated intensity along the *a*-axis and a low intensity along the *b*-axis[32,33]. Notably, we identified 13 prominent Raman peaks in the 50–300 cm$^{-1}$ spectral range, consistent with those of pristine NbIrTe$_4$[32] (**Supplementary Fig. S1**), indicating that the lattice structure of NbIrTe$_4$ remained intact throughout the entire fabrication process. Cross-sectional bright-field image of transmission electron microscopy (TEM) further confirmed the high-quality and well-defined interfaces of the multilayers (**Fig. 1d**), showing that NbIrTe$_4$ retained the single-crystalline. Additionally, energy-dispersive X-ray spectroscopy (EDX) line scans verified the deposition thickness and the presence of sharp interfaces.

**Dimensionality Evolution on Electronic Structure of NbIrTe$_4$**

The anisotropic lattice structure of NbIrTe$_4$ gives it fascinating topological properties that depend on its dimensionality. When NbIrTe$_4$ is reduced to a monolayer, translation symmetry along the c-axis is broken, leaving only identity and mirror symmetries in the b-c plane. Combined with strong spin–orbit coupling, monolayer NbIrTe$_4$ shows a global band gap and displays prominent Van Hove singularities near the Fermi level (**Fig. 2a**), suggesting a quantum spin Hall insulating (QSHI) states[34]. In contrast, the bulk of NbIrTe$_4$ hosts relativistic chiral Weyl nodes[35,36], making it a topological Weyl semimetal (TWS). To validate that, the angle-resolved photoemission spectroscopy (ARPES) measurements were performed on NbIrTe$_4$ flakes at room temperature (**Supplementary Note 2**). Along the Γ–X direction (**Fig. 2b**), second-derivative ARPES spectra reveal a blurred electronic band crossing the Fermi level near the Γ point, forming elliptical electron pockets. These ARPES results, further confirmed by density functional theory (DFT) calculations, demonstrate the distinct Weyl cone features in the bulk state of NbIrTe$_4$, underscoring its three-dimensional type-II Weyl semimetal nature [31,37-39].

The dimensional crossover in NbIrTe$_4$ significantly alters its electronic band structure, which in turn affects its spin-related topological properties. To investigate this, we calculated the OOP spin Berry curvature projections ($\Omega_{zx}^{z}$) for both monolayer and bulk NbIrTe$_4$ in the low-energy region, as shown in **Fig. 2c, d**. As the NbIrTe$_4$ evolutes from a QSH insulator in the monolayer to a TWS in the bulk, the OOP spin Berry curvature $\Omega_{zx}^{z}$ redistributes across momentum space, resulting in a more pronounced and complex spin Berry curvature landscape. This redistribution leads to significant changes in spin current generation with the dimensional crossover of NbIrTe$_4$. As shown in **Fig. 2e**, the calculated spin Hall conductivity, using the Kubo formula[40,41], reveals that this dimensional evolution enhances the OOP spin Hall conductivity ($\sigma_{sh}^{z}$) by two orders of magnitude while reducing the in-plane spin Hall conductivity ($\sigma_{sh}^{y}$), resulting in the OOP spin Hall conductivity ($\sigma_{sh}^{z}$) of bulk NbIrTe$_4$ is more than five times greater than its in-plane counterpart ($\sigma_{sh}^{y}$). Consequently, the spin tilt angle ($\varphi_{spin}^{y\text{-}z}$=arctan($\sigma_{sh}^{z}/\sigma_{sh}^{y}$)) shifts from ~5° in the monolayer to over ~80° in the bulk NbIrTe$_4$.

To elucidate the underlying dimensional evolution of electronic structure and its impact on OOP spin generation, we performed additional DFT calculations on NbIrTe$_4$ with varying layer numbers, from one to three layers. Intermediate-layer NbIrTe$_4$ demonstrates a progressive emergence of elliptical electron and hole pockets near the Γ point, culminating in crossing features reminiscent of Weyl cone as the layer number increases (**Supplementary Note3**). This electronic structure transition is accompanied by a marked enhancement in the OOP spin Hall conductivity ($\sigma_{sh}^{z}$), contrasting with the diminishing in-plane spin Hall conductivity ($\sigma_{sh}^{y}$) (**Supplementary Figure S7**).

# Thickness Dependent Spin-charge Conversion Driven by Dimensionality Crossover

Based on the above results, we confirm that the bulk TWS state of NbIrTe$_4$ plays a critical role in generating OOP spin currents. To experimentally verify this dimensional effect, we characterized the spin-charge conversion efficiency of NbIrTe$_4$ devices with varying thicknesses $t_{\text{NbIrTe4}}$ (ranging from 14 nm to 119.6 nm) using angle-dependent spin torque ferromagnetic resonance (STFMR) measurements (**Supplementary Note4**)[12,17,25]. The STFMR signal ($V_{\text{mix}}$) was decomposed into symmetric ($V_S$) and antisymmetric ($V_A$) components, corresponding to in-plane and OOP spin torques[42], respectively. The representative STFMR result of device with $t_{\text{NbIrTe4}}$ =37.6 measurement is shown in **Fig. 3a**, where the clear asymmetry in the amplitude of antisymmetric components $V_A$ under positive and negative magnetic fields, confirming the presence of $\hat{\sigma}_z$ [17]. The angular dependence of the symmetric and antisymmetric coefficients ($V_S$ and $V_A$) was further analyzed to evaluate the in-plane and OOP spin Hall conductivities (**Fig. 3b & Fig. S10**). Based on these results, the in-plane and OOP spin Hall conductivity $\sigma_{sh}^y$ and $\sigma_{sh}^z$ were determined to be 1.75 $\times$ 10$^5$ $\hbar/2e \cdot \Omega^{-1}\text{m}^{-1}$ and 0.35 $\times$ 10$^5$ $\hbar/2e \cdot \Omega^{-1}\text{m}^{-1}$, respectively, for $t_{\text{NbIrTe4}}$=14 nm. With increasing the thickness, the OOP spin Hall conductivity $\sigma_{sh}^z$ increased more than fourfold, reaching unprecedented 1.2$\times$10$^5$ $\hbar/2e \cdot \Omega^{-1}\text{m}^{-1}$, while the in-plane spin Hall conductivity $\sigma_{sh}^y$ decreased to 0.25$\times$10$^5$ $\hbar/2e \cdot \Omega^{-1}\text{m}^{-1}$ at 119.6 nm.

As shown in **Fig. 3c** the thickness-dependent trend of spin conversion efficiency aligns well with our DFT results, where $\sigma_{sh}^y$ decreases whereas $\sigma_{sh}^z$ steadily increases to saturation. Correspondingly, the spin tile angle $\varphi_{spin}^{y-z}$ increases from 11° in the atomic-thickness range to a remarkable value 77° when t$_{NbIrTe4}$ crosses a critical value of approximately 40 nm, approaching the predicted value of bulk NbIrTe$_4$ from DFT calculations. This large spin tilt angle, significantly exceeding 45°, indicates a transition from in-plane-dominant to OOP-dominant spin current, confirming that the bulk Weyl semimetal state of NbIrTe$_4$ has become the dominant.

The significant enhancement of $\sigma_{sh}^z$ provides direct experimental evidence, further underscoring the crucial contribution of the TWS nature of NbIrTe$_4$ to the OOP spin Hall conductivity. To illustrate the enhancement effect of TWS on OOP spin currents, **Fig. 3d** compares the enhanced $\sigma_{sh}^z$ and its ratio $\sigma_{sh}^z/\sigma_{sh}^y$ in thick NbIrTe$_4$ devices against previously reported spin-Hall materials (**Supplementary Table S1**) [12, 14, 16, 17, 19, 20, 23–25]. For the first time, our dimensional engineering strategy enabled the OOP spin Hall conductivity to reach the $10^5$ $\hbar/2e \cdot \Omega^{-1}m^{-1}$ regime, a milestone unattained in prior spin Hall materials. Particularly, the significant enhancement of $\sigma_{sh}^z$ in thicker devices resulted in the $\sigma_{sh}^z/\sigma_{sh}^y$ ratio exceeding value of 4, indicating a transition to spin transport dominated by OOP spin currents.

**Efficient Deterministic SOT Switching of Perpendicular Magnetization Induced by Dominant OOP Spin Currents**

The dominant OOP spin currents enable efficient and deterministic SOT switching for PMA systems. To further demonstrate this, we fabricated the Hall bar devices of NbIrTe$_4$/PMA multilayer. Anomalous Hall effect (AHE) measurements confirmed the PMA characteristics, with an anisotropy field ($\mu_0H_k$) of ~ 250 mT (**Supplementary Fig.**

**S11**). By further analyzing the effective field generated by SOT from the AHE curves, we confirmed that the OOP spin current conversion efficiency is consistent with the results obtained from STFMR measurements (**Supplementary Note 5**).

Then, we performed the room-temperature SOT switching measurement (**Methods**). The typical switching curves for the NbIrTe$_4$ device ($t_{NbIrTe4}$= 34.2 nm) under different in-plane magnetic fields was shown in **Fig. 4a**, where the field-free deterministic SOT switching was achieved (**Supplementary Note 6 for repeatability**). Notably, the SOT switching polarity remained unchanged even under external magnetic fields up to ±60 mT. This robust SOT switching polarity can be attributed to the large spin tilt angle ($\varphi_{spin}^{y\text{-}z}$) and significant OOP spin current σ$_z$[20]. The OOP spin current generates a dominant OOP torque ($\tau_z^{DL} \propto \mathbf{m} \times \hat{\sigma}_z \times \mathbf{m}$), enabling direct switching between +**m$_z$** and −**m$_z$** independent of the in-plane magnetic field (**Supplementary Note8**).

We systematically investigated room-temperature field-free SOT switching across a range of NbIrTe$_4$ thicknesses from 12 nm to 65 nm (**Fig. 4b**), where the threshold current density for deterministic switching significantly drops from 6.2 MA/cm$^2$ to an outstandingly low value of 1.8 MA/cm$^2$. This threshold current density is more than 40 times lower than that in conventional Pt-based systems[43-45], highlighting the superior energy efficiency of enhanced OOP spin current in enabling field-free switching. Additionally, the reduction in threshold current density with increasing thickness further corroborates the transition of dominant spin current induced by the dimensional crossover. Combined with macrospin simulations, we further confirmed that a larger

$\varphi_{spin}^{y-z}$ in the bulk state of NbIrTe$_4$ leads to a lower threshold current density (**Fig. 4c**), where threshold current density decreases by over 70% as $\varphi_{spin}^{y-z}$ increases from 2° to 75° (**Supplementary Note7**).

Based on above experimental and theoretical results, we conclude that increasing the thickness of NbIrTe$_4$ induces a dimensional crossover in electronic structure, transitioning from a 2D quantum spin Hall (QSH) insulator to a 3D Weyl semimetal. This dimensional transition, characterized by the emergence of relativistic chiral Weyl nodes, significantly enhances the OOP spin Hall conductivity by a factor of four, thereby effectively reducing the switching threshold current in PMA systems. Furthermore, increased thickness improves conductivity, with resistivity dropping from 235.1 μΩ·cm to 151.4 μΩ·cm, contributing to a nearly two-order-of-magnitude reduction in power consumption, down to 0.29 mW/μm³ (**Fig. 4d**). This ultra-low SOT field-free switching power is 200 times lower than that of conventional Pt-based systems, representing the lowest value among all existing SOT systems to date[7, 16, 20, 23, 27, 46–50] (**Supplementary Note9**). Combined with the strong robustness of field-free SOT switching, these breakthroughs underscore the tremendous potential of dimensionality-enhanced OOP spin current in NbIrTe$_4$ for ultralow-power and high-high-reliability spintronic technologies.

**Conclusion**

In this study, we have demonstrated the dimensionality evolution effect in enhancing out-of-plane (OOP) spin currents in NbIrTe$_4$. By tuning its dimensionality from a two-dimensional quantum spin Hall insulator to a three-dimensional type-II

Weyl semimetal, we achieved a record-breaking OOP spin Hall conductivity exceeding $10^5$ $\hbar/2e \cdot \Omega^{-1}m^{-1}$. This advancement enabled efficient and field-free SOT switching of perpendicular magnetization with ultralow threshold currents and the lowest power consumption to date. Our findings underscore the pivotal role of bulk topological Weyl semimetals in enabling robust and efficient OOP spin generation, providing a foundation for developing energy-efficient and high-density spintronic devices. Looking ahead, the dimensional engineering strategy we have established offers a versatile framework for exploring other low-symmetry quantum materials with tunable spintronic properties.

## Method

### Device Fabrication

The 2D materials NbIrTe$_4$ was pre-exfoliated from its bulk in a glove box. The exfoliated samples were immediately transferred into a high-vacuum sputter chamber with a base pressure of less than $2\times10^{-9}$ torr. Then, we use magnetron sputtering system to deposit magnetic layer, NbIrTe$_4$/Ti (2 nm)/CoFeB (1.3 nm)/MgO (2 nm)/Ta (1 nm), on the 2D flakes. The ultrathin Ti with weak SOC was used as buffer layer between the NbIrTe$_4$ ensuring minimal spin depolarization during spin transfer from the NbIrTe$_4$ layer. The CoFeB/MgO interface guarantees PMA, while the CoFeB thickness (1.3 nm) allows complete absorption of spin current without reflection due to its short dephasing length After depositing the ferromagnetic film, we use angle dependent Ramn measurement to confirm the crystallography and atomic force microscopy to measure the thickness of the NbIrTe$_4$. Then, the samples were fabricated into Hall devices using a standard electron beam lithography.

### Raman Measurement

A linearly polarized 532 nm laser was focused onto the samples using a ×100 objective lens. The scattered Raman signals went through the same lens and back to a spectrometer in the parallel configuration. The samples were rotated by a stage with a step size of 6° and all the other parameters are maintained to be identical during the measurements.

### ARPES Measurements

ARPES measurements were performed at room temperature using He–I (21.218 eV) photons and a SPECS PHOIBOS150 hemispherical energy analyzer. The base pressure of the analyzer chamber is $2 \times 10^{-10}$ mbar. The energy resolution is less than 35 meV and the angular resolution is 0.05°. A single crystal sample of NbIrTe$_4$ is cleaved *in situ* in the sample preparation chamber with a base pressure of $3 \times 10^{-8}$ mbar to obtain clean surface and then transferred to the analyzer chamber.

### Electric Measurements

The electric measurements were performed using a combination of Keithley 6221 and 2182A. For switching measurement, Current pulses with a duration of 100 μs were applied along the low-symmetry a-axis of NbIrTe$_4$. After each write current pulse, a small read current pulse of 0.1 mA was applied to detect the magnetization state via the anomalous Hall resistance.

**STFMR measurement**

We first performed STFMR measurements with a fixed magnetic field angle of $\phi_H \sim 40°$, where an RF current ($I_{rf}$) is applied along the a-axis at a frequency of 6 GHz. Then, we performed angle-dependent STFMR measurements by varying the magnetic field angle $\phi_H$. In general, the spin current that generates torque has components along all three $x$, $y$ and $z$ axes, thus angular dependencies for the coefficients S and A are:

$$S = S_{DL}^X \sin\phi \sin 2\phi + S_{DL}^Y \cos\phi \sin 2\phi + S_{FL}^Z \sin 2\phi$$
$$A = A_{FL}^X \sin\phi \sin 2\phi + A_{FL}^Y \cos\phi \sin 2\phi + A_{DL}^Z \sin 2\phi \qquad (1)$$

where $S_{DL}^X$, $S_{DL}^Y$, and $A_{DL}^Z$ is the coefficients for the damping-like torque for $x, y, z$ polarized spin current; $A_{FL}^X$ $A_{FL}^Y$ and $S_{FL}^Z$ are the corresponding field-like torques for $x, y, z$ polarized spin current respectively.

**First Principles Calculations**

We use the first-principles theory to calculate the electronic structures and spin Hall conductivity. The electronic structures calculations were performed using the Vienna Ab-initio Simulation Package (VASP). The Perdew–Burke–Ernzerhof (PBE) parameterizations were used for the generalized gradient approximation (GGA). The projector augmented-wave method was used with an energy cut-off of 500 eV. Spin–orbit coupling was included. The bulk NbIrTe$_4$ with $k$-point sampling grids of 13 × 6 × 6 were used. For the calculations of the spin Hall conductivity, we firstly use the Wannier90 code to construct the tight-binding Hamiltonian using Nb-d, Nb-p, Ir-d, and Te-p-derived bands. Then the spin Hall conductivity is calculated by Wannier-berri code using he Kubo formula:

$$\sigma_{\alpha\beta}^{\text{SHC},\gamma} = \frac{-e\hbar}{N_k V_c} \sum_{\mathbf{k}} \sum_{n,m} (f_{n\mathbf{k}} - f_{m\mathbf{k}}) \frac{\text{Im}\left[\langle\psi_{n\mathbf{k}}|\frac{1}{2}\{s^\gamma, v_\alpha\}|\psi_{m\mathbf{k}}\rangle\langle\psi_{m\mathbf{k}}|v_\beta|\psi_{n\mathbf{k}}\rangle\right]}{(\varepsilon_{n\mathbf{k}} - \varepsilon_{m\mathbf{k}})^2 - (\hbar\omega + i\eta)^2} \qquad (2)$$

where $s^\gamma$ and $v_\alpha$ ($\alpha, \beta, \gamma = x, y, z$) are spin and velocity operators respectively. $\hat{j}_\gamma^\alpha = \frac{1}{2}\{s^\gamma, v_\alpha\}$ is the spin current operator for a spin current polarized along $\gamma$ and transport along α direction, $f_{n\mathbf{k}}$ is the Fermi–Dirac distribution function for band $n$ and wave vector $\mathbf{k}$. In the d.c. limit ($\omega = \eta = 0$).

**Macrospin Simulations**

The macrospin simulations were performed based on the Landau–Lifshitz–Gilbert equation, where the spin current was set as $\sigma = (0, -\cos\varphi_{spin}^{y-z}, \sin\varphi_{spin}^{y-z})$. The magnetic parameters, such as the saturation magnetization $M_s = 8 \times 10^5$ A m$^{-1}$ and anisotropy constant $K_u = 9 \times 10^5$ J m$^{-3}$, used for the simulations are obtained from the experiments. A Gilbert damping constant $\alpha = 0.01$ is used for CoFeB. The effective spin Hall angle $\theta_{sh} = 0.2$ and variable tilt angle $\varphi_{spin}^{y-z}$ are used to convert the parameters into transverse spin currents. The spin dynamic simulations are performed from 0 to 4.5 ns with a time interval of 10 ps.

ACKNOWLEDGMENT

This work was supported in part by the National Natural Science Foundation of China (Nos. 52261145694, T2394475, 62371019, 62174010, and 92164206), the Beijing Natural Science Foundation (Nos. 4232070 and 4222070), S&T Project(Z221100007722030), the International Mobility Project (No. B16001), ERC CoG project MAGNETALLIEN grant ID 101086807, "Lorraine Université d'Excellence" reference ANR-15-IDEX-04-LUE and the China Scholarship Council (CSC). The authors also thank CC MiNaLor of IJL and Nanofabrication facility of Beihang Nano for device fabrication.

AUTHOR CONTRIBUTIONS

W.Y., J.C.R.S. and X.Y.L. conceived the original idea; W.Y., X.Y.L., W.S.Z. and J.C.R.S. planned and designed the experiments; W.Y. and J.N.L. fabricated the samples with help of M.H., and S.M.; W.Y, J.N.L, X.H.W. and J.C.R.S. performed STFMR and d.c SOT measurements with help of H.D. and S.P-W; W.Y. performed the Raman measurement; X.Y.W. and X.Y.L. performed the ARPES measurements; W.Y. performed the DFT calculation and macrospin simulations; W.Y, X.H.W., J.N.L., X.Y.L, S.P-W., S.M., W.S.Z., J.C.R.S., A.F discussed and analyzed the data; W.Y., X.Y.L., J.C.R.S., A.F. and W.S.Z. wrote the manuscript with help from all the authors.

COMPETING FINANCIAL INTERESTS STATEMENT

The authors declare no competing financial interests.

**Figures**

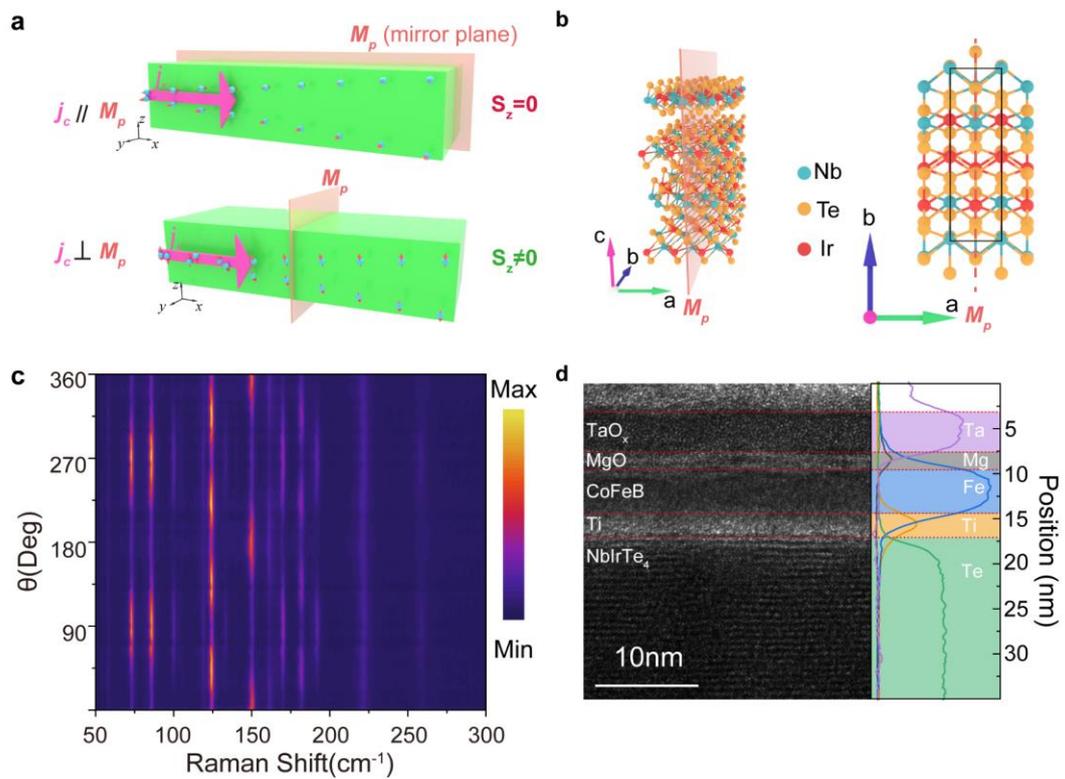

**Fig. 1 | Symmetry and Lattice of NbIrTe$_4$ for Out-of-Plane (OOP) Spin Current Generation. a,** Symmetry analysis for the generation of unconventional OOP spin currents in NbIrTe$_4$. **b,** Schematic crystal structure of NbIrTe$_4$, illustrating the layered arrangement and mirror symmetry in the *b-c* plane. **c,** Angle-dependent Raman measurement of NbIrTe$_4$. **d**, HR-TEM image and EDX line profile of the as-deposited NbIrTe$_4$/Ti/CoFeB/MgO/TaO$_x$, showing good crystal quality of NbIrTe$_4$.

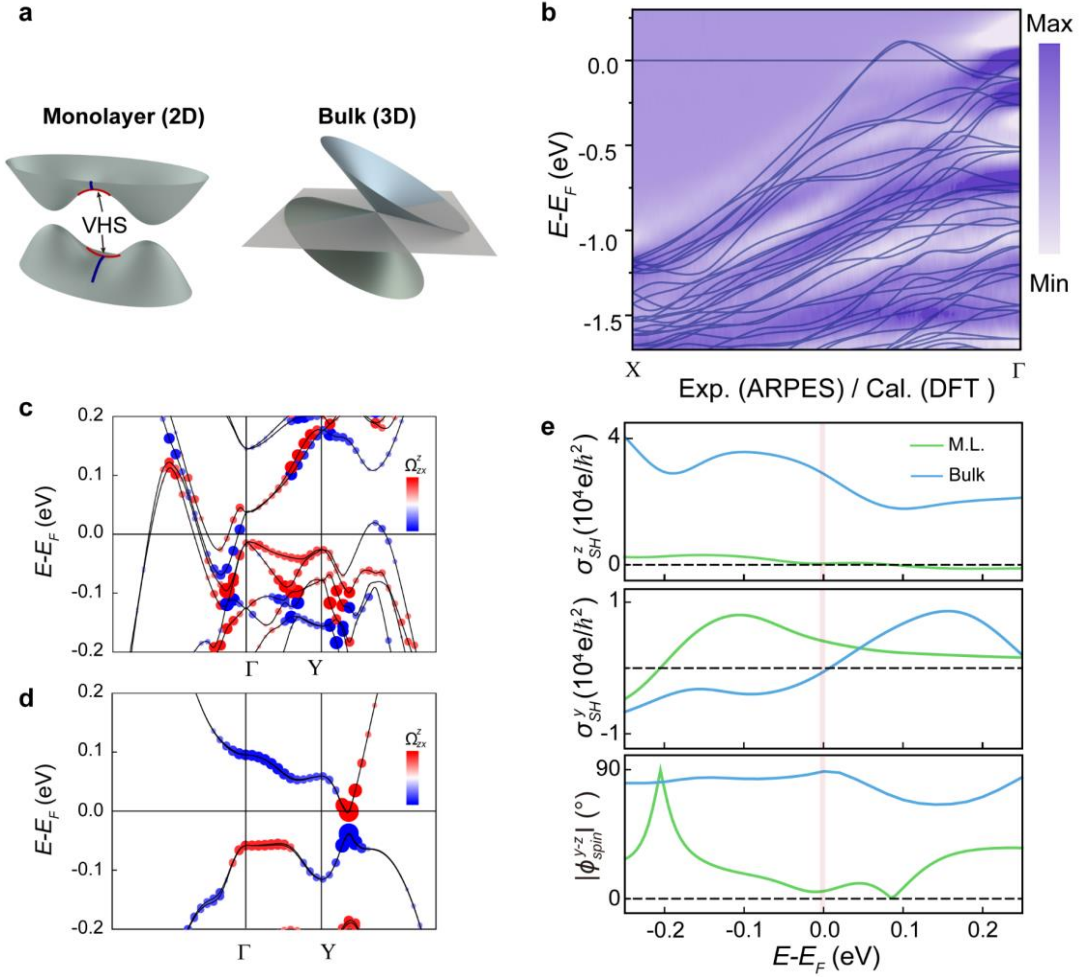

**Fig. 2 | Dimensionality Evolution Enhanced OOP Spin Current Generation of NbIrTe$_4$. a**, A sketch of the parabolic band structure with van Hove singularities (VHSs) and tilted Weyl cones for monolayer and bulk NbIrTe$_4$. **b,** Spin-resolved angle-resolved photoemission spectroscopy (ARPES) results compared with density functional theory (DFT) band structure of bulk NbIrTe$_4$. The band structure highlights the prominent bulk contributions and the presence of elliptical electron pockets and Weyl points near the Fermi level. **c & d,** DFT-calculated OOP spin Berry curvature projections ($\Omega_{zx}^z$) for monolayer (**c**) and bulk (**d**) NbIrTe$_4$ in the low-energy region. **e,** DFT-calculated spin Hall conductivity ($\sigma_{sh}^{z(y)}$) and absolute spin tilt angle $|\varphi_{spin}^{y-z}|$ of monolayer and bulk NbIrTe$_4$.

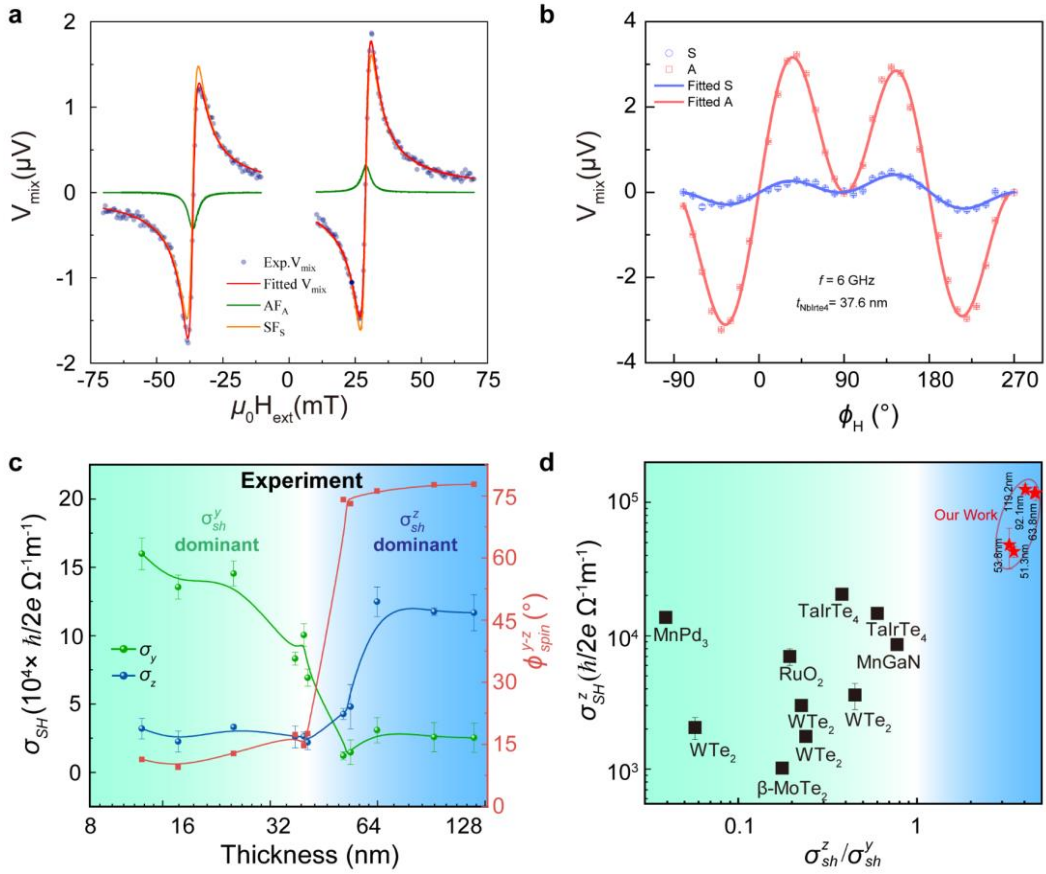

**Fig. 3 | Dimensionality Transition of Spin-Charge Conversion Efficiency in NbIrTe4. a,** Spin torque ferromagnetic resonance (STFMR) measurements of device with $t_{NbIrTe4}$=37.6nm. The RF current of 6 GHz was applied along the a-axis. The green and yellow lines are antisymmetric ($AF_A$) and symmetric contributions ($SF_S$) in fitting curve, respectively. **b,** Angle-dependent symmetric (S) and antisymmetric (A) component extracted from STFMR measurements obtained by varying $\phi_H$, the red and blue line is fitted based on Eq. 2 for A and S components. The results yields $\theta_{sh,y}$ = 0.26±0.02, $\theta_{sh,y}$ = 0.02±0.01 and $\theta_{sh,z}$ = 0.08±0.02. **c,** Spin Hall conductivity ($\sigma_{sh}^{z(y)}$) and $\varphi_{spin}^{y-z}$ from angle-dependent STFMR measurements in NbIrTe4 devices of varying thickness. **d**, Comparison of OOP spin Hall conductivity $\sigma_{sh}^z$ and the ratio $\sigma_{sh}^z/\sigma_{sh}^y$ with previously reported works[12,14,16,17,19-21,23-25].

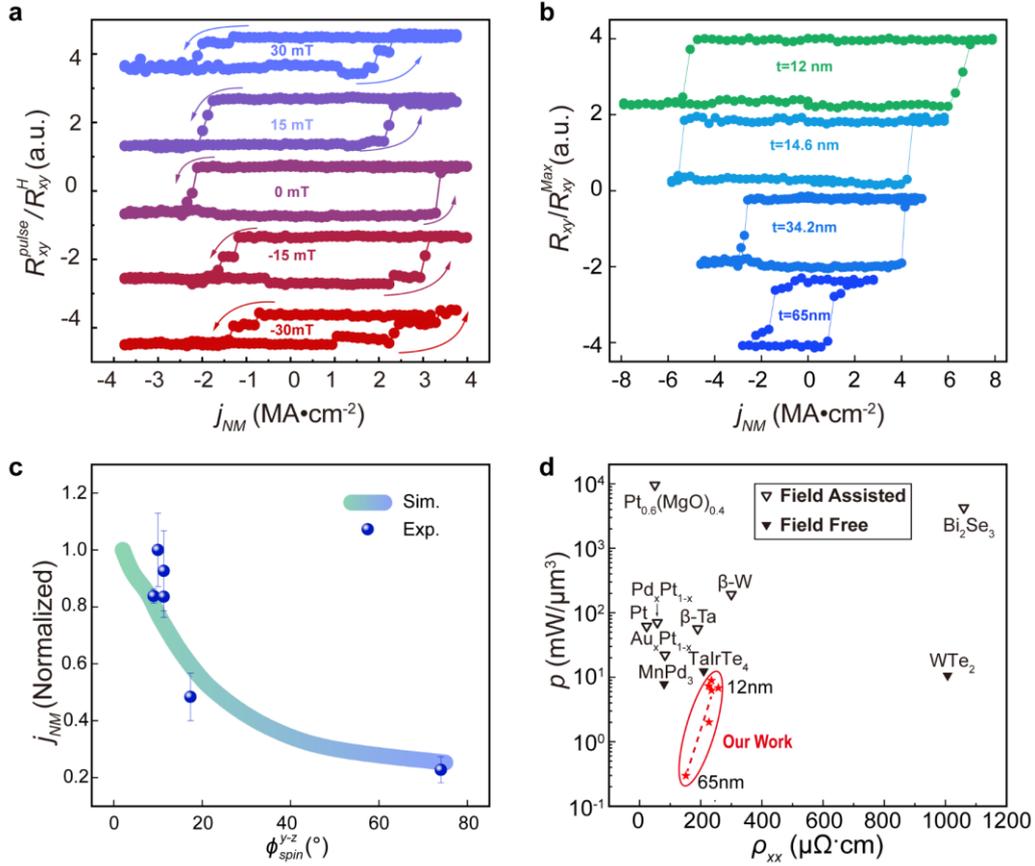

**Fig. 4 | Dimensionality-Enhanced OOP Spin Currents of NbIrTe₄ Enable Ultralow-Power SOT Switching. a,** Current induced magnetization switching under different in-plane field ($H_x$). The invariance of the SOT switching polarity under an in-plane field confirms the robustness of the OOP SOT in NbIrTe₄-based devices. **b,** Room-temperature and field-free SOT switching in devices with varying NbIrTe₄ thicknesses. As the thickness increases, the threshold current density for field-free SOT switching significantly decreases, reaching as low as 1.8 MA/cm². **c,** Relationship between spin tilt angle ($\varphi_{spin}^{y-z}$) and critical current density ($j_{NM}$) for field free SOT switching. The blue points represent experimental data, where $j_{NM}$ is extracted from SOT switching measurements, and spin tilt angle $\varphi_{spin}^{y-z}$ is determined from angle-dependent STFMR results. The solid line represents micromagnetic simulation results.

**d**, Benchmarking the power consumption ($p$) required for SOT-induced switching in PMA systems as a function of the longitudinal resistance $\rho_{xx}$ for various spin sources[6, 7, 16, 20, 23, 27, 46–50].

# References:

1   Dieny, B. *et al.*, Opportunities and challenges for spintronics in the microelectronics industry. *NAT ELECTRON* **3** 446 (2020).

2   Manchon, A. *et al.*, Current-induced spin-orbit torques in ferromagnetic and antiferromagnetic systems. *REV MOD PHYS* **91** 35004 (2019).

3   Yang, H. *et al.*, Two-dimensional materials prospects for non-volatile spintronic memories. *NATURE* **606** 663 (2022).

4   Zongxia, G. *et al.*, Spintronics for Energy- Efficient Computing: An Overview and Outlook. *P IEEE* **109** 1398 (2021).

5   Lin, X., Yang, W., Wang, K. & Zhao, W., Two-dimensional spintronics for low-power electronics. *NAT ELECTRON* **2** 274 (2019).

6   Miron, I. M. *et al.*, Perpendicular switching of a single ferromagnetic layer induced by in-plane current injection. *NATURE* **476** 189 (2011).

7   Liu, L. *et al.*, Spin-Torque Switching with the Giant Spin Hall Effect of Tantalum. *SCIENCE* **336** 555 (2012).

8   Liu, L., Lee, O. J., Gudmundsen, T. J., Ralph, D. C. & Buhrman, R. A., Current-Induced Switching of Perpendicularly Magnetized Magnetic Layers Using Spin Torque from the Spin Hall Effect. *PHYS REV LETT* **109** 96602 (2012).

9   Wu, H. *et al.*, Field-free approaches for deterministic spin-orbit torque switching of the perpendicular magnet. *Materials Futures* **1** 22201 (2022).

10  Kao, I. *et al.*, Deterministic switching of a perpendicularly polarized magnet using unconventional spin–orbit torques in $WTe_2$. *NAT MATER* **9** 1029 (2022).

11  Baek, S. C. *et al.*, Spin currents and spin–orbit torques in ferromagnetic trilayers. *NAT MATER* **17** 509 (2018).

12  Zhang, Y. *et al.*, Room temperature field-free switching of perpendicular magnetization through spin-orbit torque originating from low-symmetry type II Weyl semimetal. *SCI ADV* **9** eadg9819 (2023).

13  Li, Z. *et al.*, Collinear Spin Current Induced by Artificial Modulation of Interfacial

Symmetry. *ADV SCI* **11** 2406924 (2024).

[14] Wang, F. *et al.*, Field-free switching of perpendicular magnetization by two-dimensional PtTe$_2$/WTe$_2$ van der Waals heterostructures with high spin Hall conductivity. *NAT MATER* **23** 768 (2024).

[15] Liu, Y. & Shao, Q., Two-Dimensional Materials for Energy-Efficient Spin–Orbit Torque Devices. *ACS NANO* **14** 9389 (2020).

[16] Wang, X. *et al.*, Room Temperature Field-free Switching of CoFeB/MgO heterostructure based on large-scale few-layer WTe$_2$. *CELL REP PHYS SCI* **4** 101468 (2023).

[17] MacNeill, D. *et al.*, Control of spin–orbit torques through crystal symmetry in WTe$_2$/ferromagnet bilayers. *NAT PHYS* **13** 300 (2016).

[18] Shin, I. *et al.*, Spin–orbit torque switching in an all-van der Waals heterostructure. *ADV MATER* **34** 2101730 (2022).

[19] Shi, S. *et al.*, Observation of the out-of-plane polarized spin current from CVD grown WTe$_2$. *ADV QUANTUM TECHNOL* **4** 2100038 (2021).

[20] Liu, Y. *et al.*, Field-free switching of perpendicular magnetization at room temperature using out-of-plane spins from TaIrTe$_4$. *NAT ELECTRON* **6** 732 (2023).

[21] Li, R. *et al.*, Layer-dependent spin-orbit torques generated by the centrosymmetric transition metal dichalcogenide β−MoTe$_2$. *PHYS REV B* **100** 184402 (2019).

[22] Chen, X. *et al.*, Observation of the antiferromagnetic spin Hall effect. *NAT MATER* **20** 800 (2021).

[23] DC, M. *et al.*, Observation of anti-damping spin–orbit torques generated by in-plane and out-of-plane spin polarizations in MnPd$_3$. *NAT MATER* **22** 591 (2023).

[24] Nan, T. *et al.*, Controlling spin current polarization through non-collinear antiferromagnetism. *NAT COMMUN* **11** 4671 (2020).

[25] Bose, A. *et al.*, Tilted spin current generated by the collinear antiferromagnet ruthenium dioxide. *NAT ELECTRON* **5** 267 (2022).

[26] Mellnik, A. R. *et al.*, Spin-transfer torque generated by a topological insulator. *NATURE* **511** 449 (2014).

[27] Mahendra, D. *et al.*, Room-temperature high spin–orbit torque due to quantum

confinement in sputtered Bi$_x$Se$_{(1-x)}$ films. *NAT MATER* **17** 800 (2018).

28  Zhang, Z. *et al.*, Field-Free Spin-Orbit Torque Switching in Perpendicularly Magnetized Ta/CoFeB/MgO/NiO/Ta with a Canted Antiferromagnetic Insulator NiO Interlayer. *ADV FUNCT MATER* 2414643 (2024).

29  Wang, X. *et al.*, Spin manipulation by giant valley-Zeeman spin-orbit field in atom-thick WSe$_2$. *APPL PHYS REV* **9** 31402 (2022).

30  Mar, A. & Ibers, J. A., Synthesis and physical properties of the new layered ternary tellurides MIrTe$_4$ (M= Nb, Ta), and the structure of NbIrTe4. *J SOLID STATE CHEM* **97** 366 (1992).

31  Mar, A., Jobic, S. & Ibers, J. A., Metal-metal vs tellurium-tellurium bonding in WTe$_2$ and its ternary variants TaIrTe$_4$ and NbIrTe$_4$. *J AM CHEM SOC* **114** 8963 (1992).

32  Zhang, J. *et al.*, Colossal Room-Temperature Terahertz Topological Response in Type-II Weyl Semimetal NbIrTe$_4$. *ADV MATER* **34** 2204621 (2022).

33  Shojaei, I. A. *et al.*, A Raman probe of phonons and electron–phonon interactions in the Weyl semimetal NbIrTe$_4$. *SCI REP-UK* **11** 8155 (2021).

34  Tang, J. *et al.*, Dual quantum spin Hall insulator by density-tuned correlations in TaIrTe$_4$. *NATURE* **628** 515 (2024).

35  Armitage, N. P., Mele, E. J. & Vishwanath, A., Weyl and Dirac semimetals in three-dimensional solids. *REV MOD PHYS* **90** 15001 (2018).

36  Manna, K., Sun, Y., Muechler, L., Kübler, J. & Felser, C., Heusler, Weyl and Berry. *NAT REV MATER* **3** 244 (2018).

37  Lee, J. *et al.*, Spin-orbit-splitting-driven nonlinear Hall effect in NbIrTe$_4$. *NAT COMMUN* **15** 3971 (2024).

38  Zhou, W. *et al.*, Nonsaturating Magnetoresistance and Nontrivial Band Topology of Type-II Weyl Semimetal NbIrTe$_4$. *ADV ELECTRON MATER* **5** 1900250 (2019).

39  Ekahana, S. A. *et al.*, Topological Lifshitz transition of the intersurface Fermi-arc loop in NbIrTe$_4$. *PHYS REV B* **102** 85126 (2020).

40  Qiao, J., Zhou, J., Yuan, Z. & Zhao, W., Calculation of intrinsic spin Hall conductivity by Wannier interpolation. *PHYS REV B* **98** 214402 (2018).

41  Tsirkin, S. S., High performance Wannier interpolation of Berry curvature and related quantities with WannierBerri code. *NPJ COMPUT MATER* **7** 33 (2021).

42  Liu, L., Moriyama, T., Ralph, D. C. & Buhrman, R. A., Spin-torque ferromagnetic resonance induced by the spin Hall effect. *PHYS REV LETT* **106** 36601 (2011).

43  Yu, G. *et al.*, Switching of perpendicular magnetization by spin–orbit torques in the absence of external magnetic fields. *NAT NANOTECHNOL* **9** 548 (2014).

44  van den Brink, A. *et al.*, Field-free magnetization reversal by spin-Hall effect and exchange bias. *NAT COMMUN* **7** (2016).

45  Pham, T. H. *et al.*, Thermal contribution to the spin-orbit torque in metallic-ferrimagnetic systems. *PHYS REV APPL* **9** 64032 (2018).

46  Zhu, L. *et al.*, Strong Damping-Like Spin-Orbit Torque and Tunable Dzyaloshinskii–Moriya Interaction Generated by Low-Resistivity $Pd_{1-x}Pt_x$ Alloys. *ADV FUNCT MATER* **29** 1805822 (2019).

47  Pai, C. *et al.*, Spin transfer torque devices utilizing the giant spin Hall effect of tungsten. *APPL PHYS LETT* **101** 122404 (2012).

48  Zhu, L., Ralph, D. C. & Buhrman, R. A., Highly Efficient Spin-Current Generation by the Spin Hall Effect in $Au_{1-x}Pt_x$. *PHYS REV APPL* **10** 31001 (2018).

49  Zhu, L., Zhu, L., Sui, M., Ralph, D. C. & Buhrman, R. A., Variation of the giant intrinsic spin Hall conductivity of Pt with carrier lifetime. *SCI ADV* **5** eaav8025 (2019).

50  Han, J. *et al.*, Room-temperature spin-orbit torque switching induced by a topological insulator. *PHYS REV LETT* **119** 77702 (2017).

# Supplementary Information

# Table of Content

**Supplementary Figures**

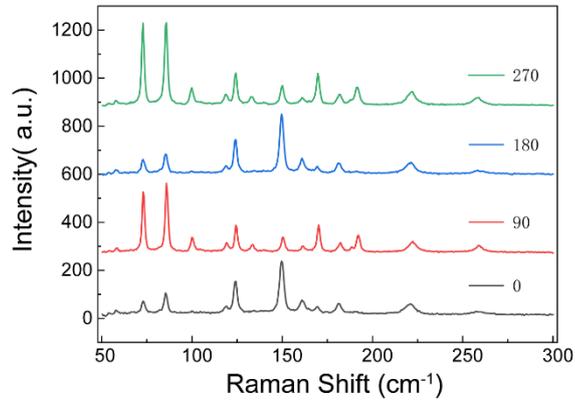

**Fig. S1 | The Representative Raman signal for varying angle between laser polarization and a-axis of NbIrTe$_4$.** Raman spectroscopy of NbIrTe$_4$ nanosheets showed 13 notable Raman peaks ranging from 50 to 300 cm$^{-1}$, which belong to A1 or A2 Raman vibrational modes.

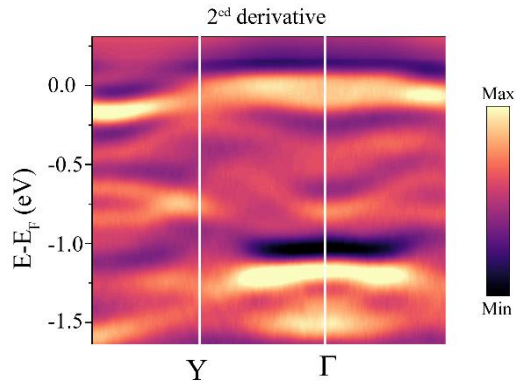

**Fig. S2 | Second-derivative ARPES spectra along the Γ–Y direction for NbIrTe$_4$.** These data complement the Γ–X measurements presented in the main text and further validate the overall consistency between experiment and first-principles calculations

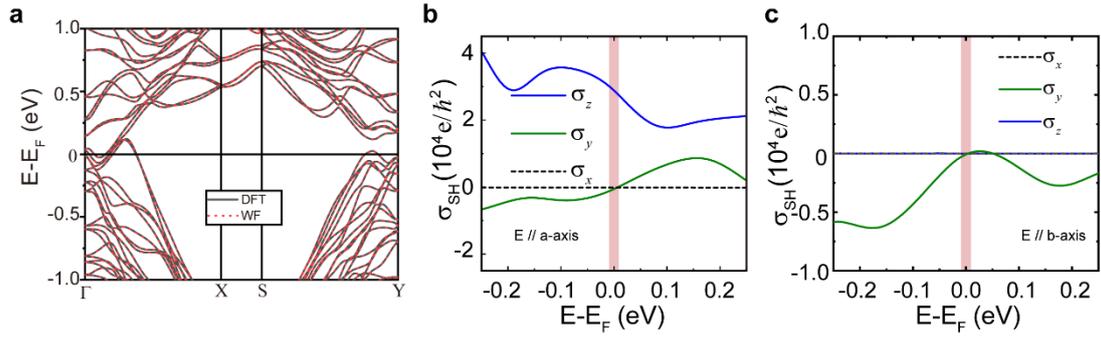

**Fig. S3 | a,** Comparison of the DFT-calculated band structure (black) and its Wannier function (WF) interpolation (red) for bulk NbIrTe$_4$, illustrating the consistency between ab initio and WF-based electronic structure descriptions. **b, c,** Spin Hall conductivity calculated along the a-axis (**b**) and b-axis (**c**) of bulk NbIrTe$_4$.

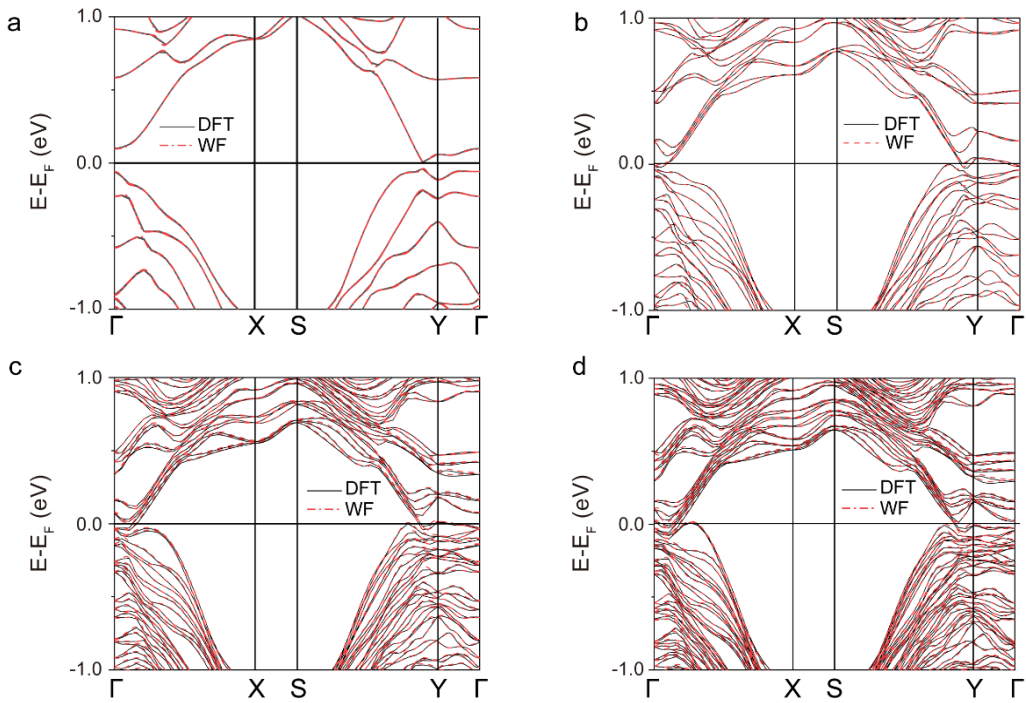

**Figure S 4 Comparison of the DFT-calculated band structure (black lines) and its Wannier function (WF) interpolation (red lines) for NbIrTe$_4$ with varying layer numbers. a**, 1 monolayer (M.L.) **b**, 2 M.L. **c**, 3 M.L. **d**, 4 M.L.

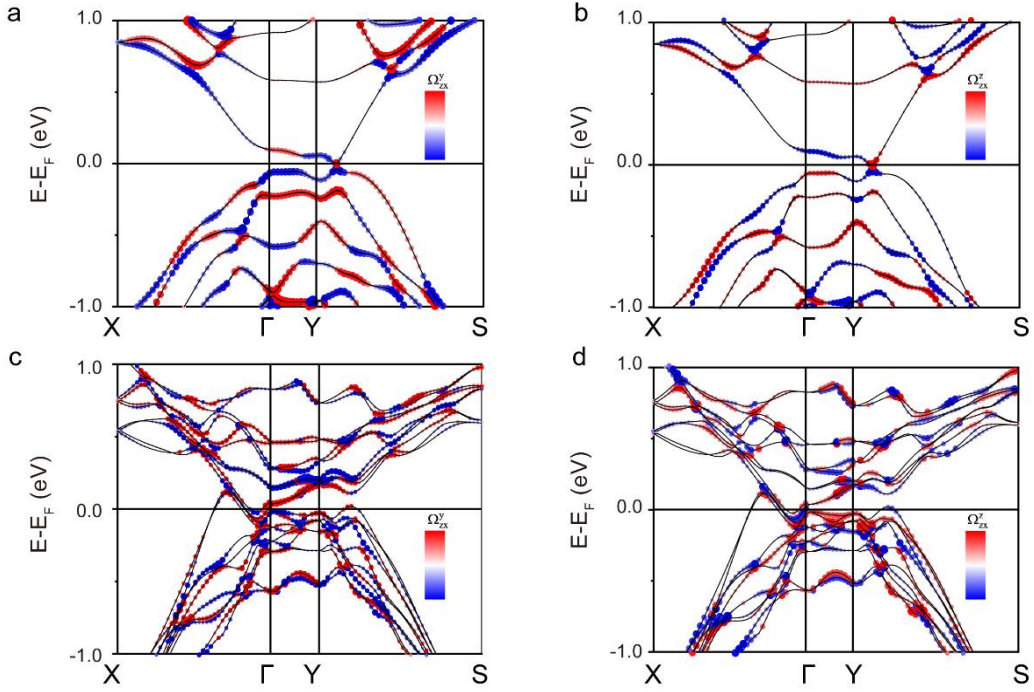

**Fig. S5 | Band-projected spin Berry curvature for single-layer and bulk NbIrTe4.**
**a, b,** the spin Berry curvature $\Omega_{zx}^{y}$ (a) and $\Omega_{zx}^{z}$ (b) for monolayer NbIrTe4. **c, d,** the spin Berry curvature $\Omega_{zx}^{y}$ (c) and $\Omega_{zx}^{z}$ (d) for bulk NbIrTe4.

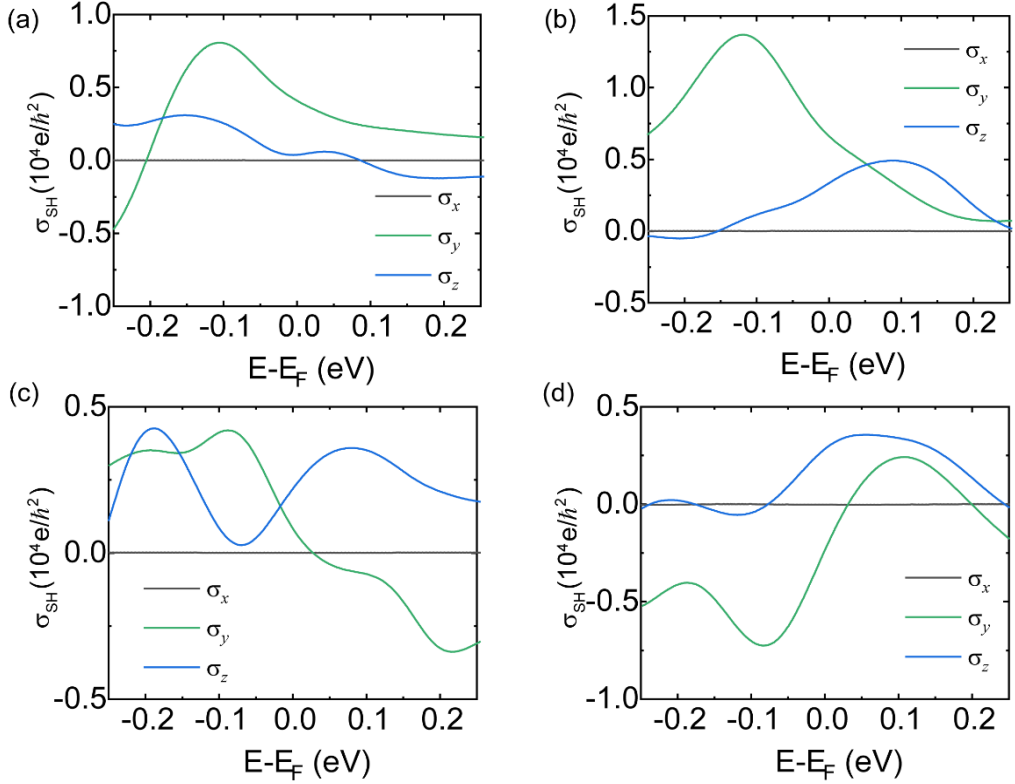

**Figure S 6 Layer dependent spin hall conductance for NbIrTe4 with varying layer numbers. a,** 1 monolayer (M.L.) **b,** 2 M.L. **c,** 3 M.L. **d,** 4 M.L.

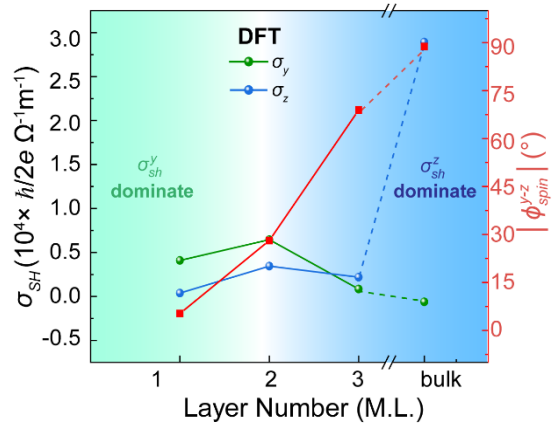

**Fig. S 7 | Calculated spin Hall conductance components ($\sigma_{sh}^{z(y)}$) and $\varphi_{spin}^{y-z}$ at the Fermi level for few-layer NbIrTe4.** The DFT results illustrating the transition from in-plane to out-of-plane dominated spin currents as thickness increases

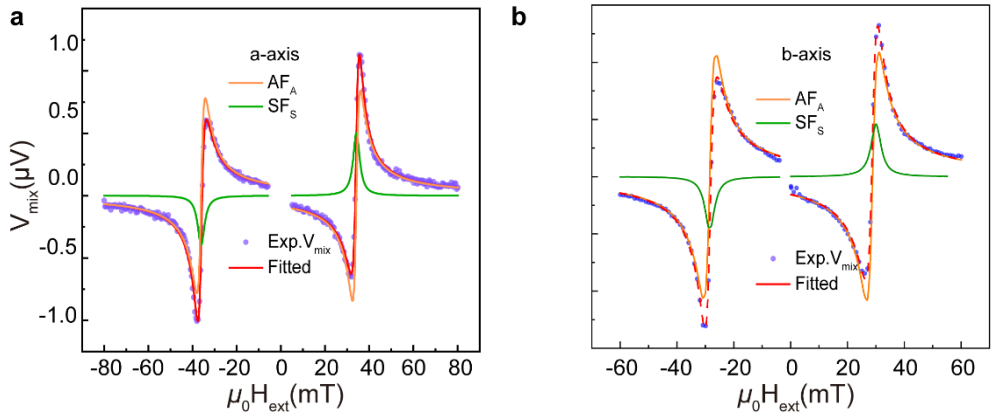

**Fig. S8 | Typical STFMR data at the frequency of 6 GHz. a,b,** ST-FMR data when $I_{rf}$ is applied along *a*-axis (**a**) and b-axis (**b**) The magnetic field is applied at fixed angle $\phi$ = 40° ($\phi$ = 220°) for positive (negative) respectively.

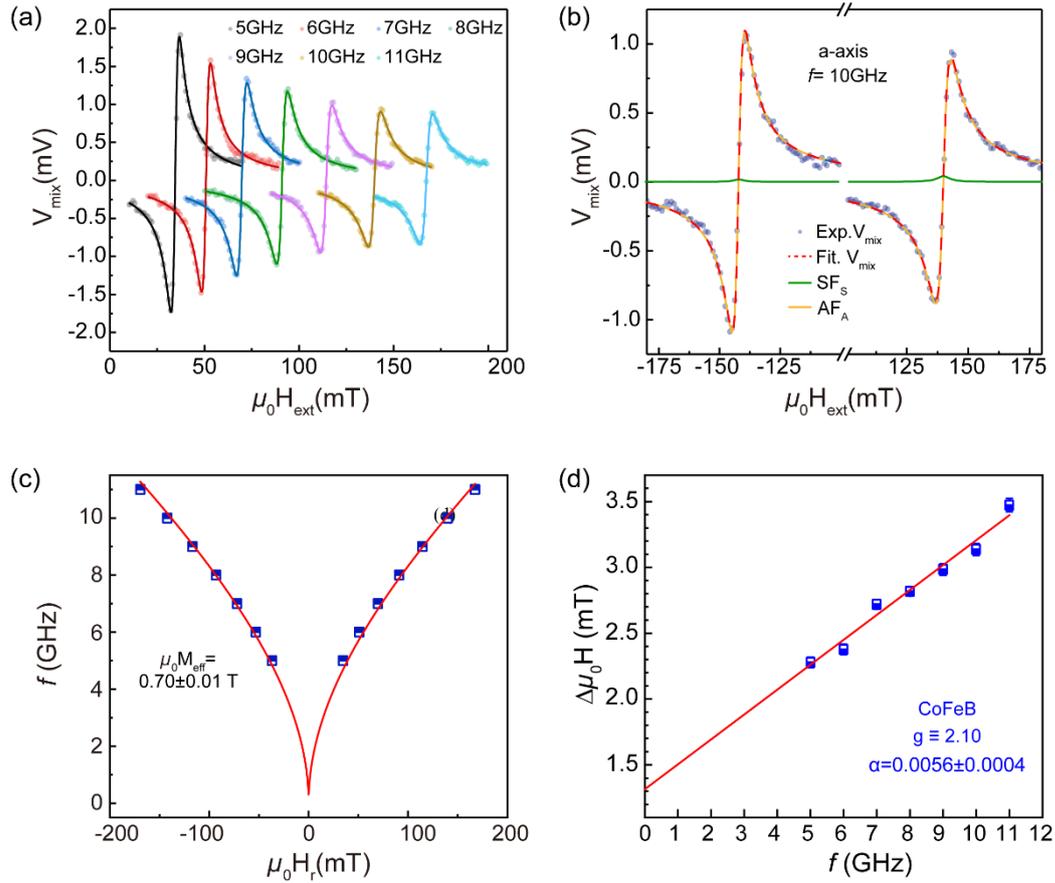

**Fig. S9 | STFMR measurement results in NbIrTe4(119.2)/CoFeB(7) device with in-plane magnetic field at fixed angle $\phi_H$ = 40°. a,** Field-dependent ST-FMR signals measured at frequencies ranging from 6 to 11 GHz. The solid lines are fits based on Eq. S4. **b,** ST-FMR signal at $f$=10 GHz under positive and negative magnetic fields ($\pm\mu_0H_{ext}$). The solid lines are also fits using Eq. S4, with the yellow and green lines representing the antisymmetric ($V_A$) and symmetric ($V_S$) components, respectively. The noticeable change in the $V_A$ component between $+\mu_0H_{ext}$ and $-\mu_0H_{ext}$ indicates the OOP spin current contribution. **c,** Plot of the resonance frequency $f$ versus resonance field $\mu_0H_r$. From this data, the effective magnetization $\mu_0M_{eff}$ was estimated as 0.7 $\pm$0.01 T. **d,** The linewidth $\Delta\mu_0H$ versus frequency. The Gilbert damping constant $\alpha$ of CoFeB was estimated as 0.0056 $\pm$0.0004.

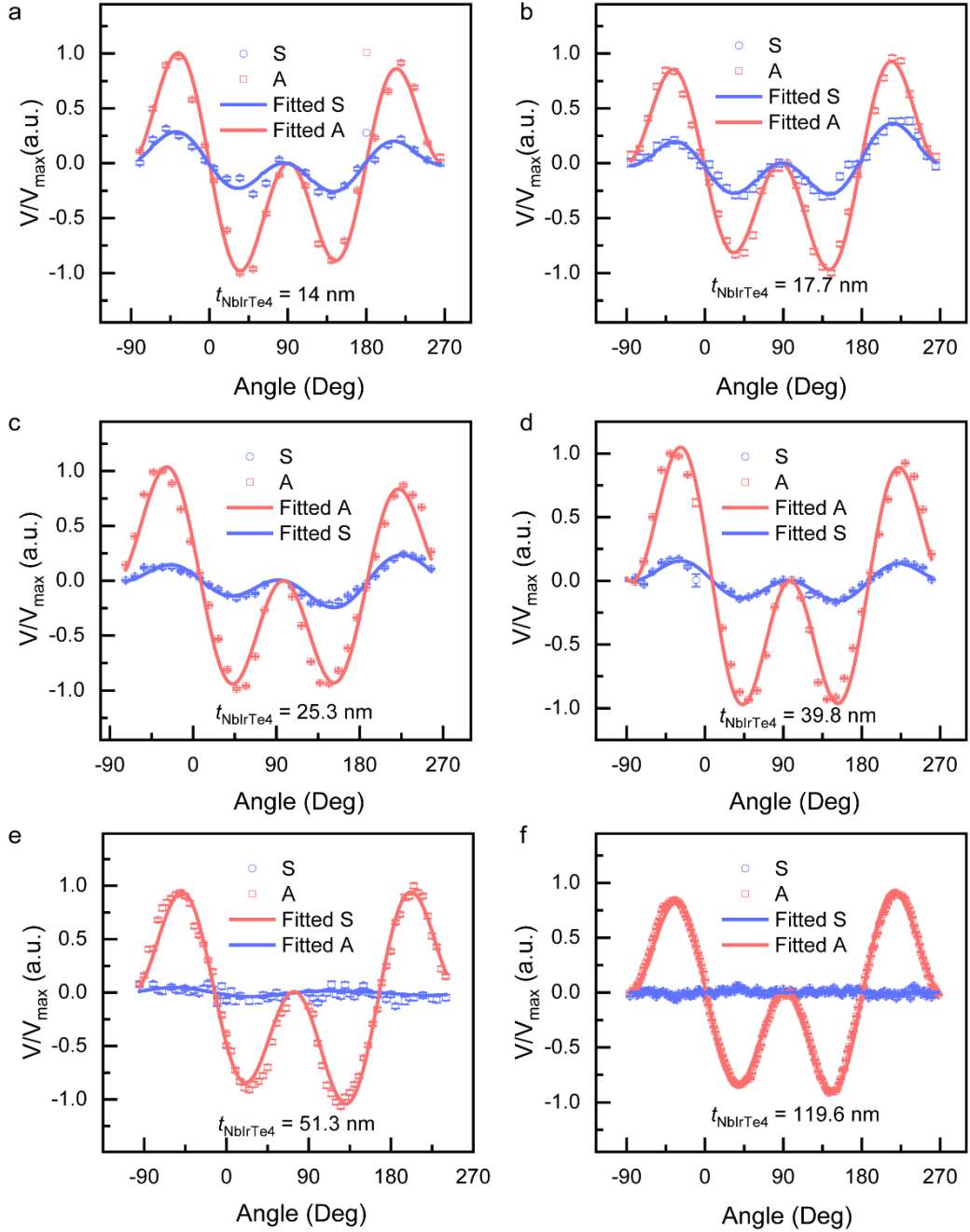

**Fig. S10 | Angle-dependent analysis of the antisymmetric ($V_A$) and symmetric ($V_S$) components for devices with $t_{NbIrTe4}$ ranging from 14 to 119.6 nm. a**, $t_{NbIrTe4}$=14 nm. **b**, $t_{NbIrTe4}$=17.7 nm. **c**, $t_{NbIrTe4}$=25.3nm. **c**, $t_{NbIrTe4}$=37.6nm. **d**, $t_{NbIrTe4}$=39.8nm. **e**, $t_{NbIrTe4}$=51.3nm. **f,** $t_{NbIrTe4}$=119.6nm. Both $V_A$ and $V_S$ were extracted from the ST-FMR signals using Eq. S4.

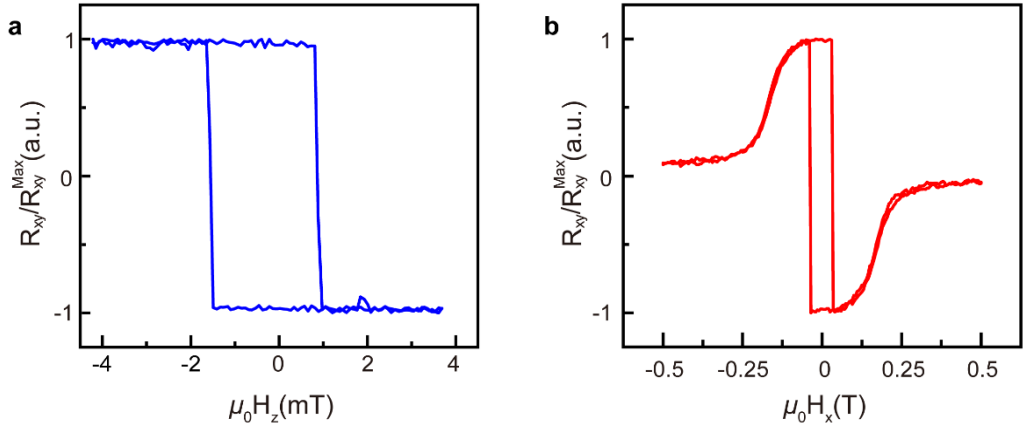

**Figure S11 | Perpendicular magnetic anisotropy of NbIrTe₄ Hall device. a**, Anomalous Hall effect (AHE) loop obtained by sweeping the OOP field ($\mu_0H_z$) under a DC current of 0.1 mA. **b**, AHE loop obtained by sweeping the in-plane field ($\mu_0H_x$) with the same DC current. The square-shaped loop and extracted anisotropy field ($\mu_0H_a$~250 mT) confirm the strong PMA in the device. The AHE signal in both (b) and (c) is normalized for clarity.

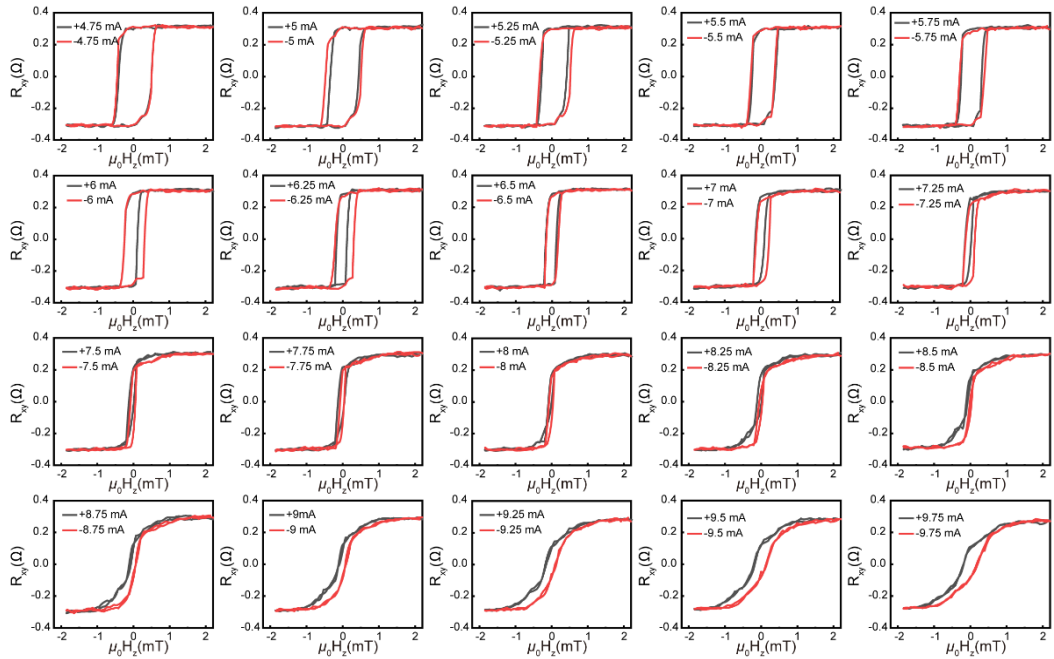

**Fig. S12 | Current dependent AHE measurement on NbIrTe₄ ($t_{NbIrTe4}$=34.2 nm)-based Hall devices.** As the applied current is varied from 4.75 mA to 9.75 mA, the AHE loops remain unchanged below a threshold current. Once the current exceeds threshold value, the AHE loops shift to the right (positive current) or left (negative

current), indicating the presence of an out-of-plane (OOP) spin current that exerts a torque on the perpendicular magnetization.

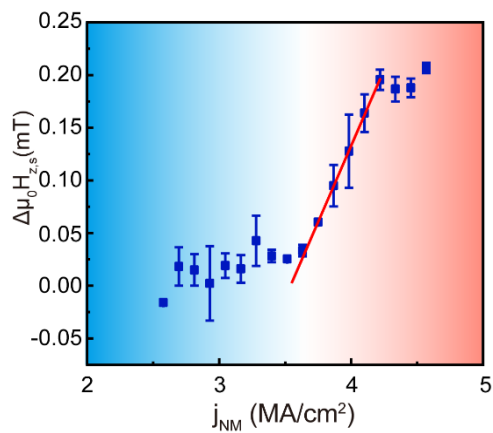

**Fig. S13 | The extracted OOP effective field ($\Delta\mu_0 H_{z,s}$) from AHE measurements as a function of current density**. Once the current density surpasses a critical threshold, $\Delta\mu_0 H_{z,s}$ increases linearly and eventually saturates. The OOP spin hall angle was determined as $0.06\pm0.002$ from the linear region.

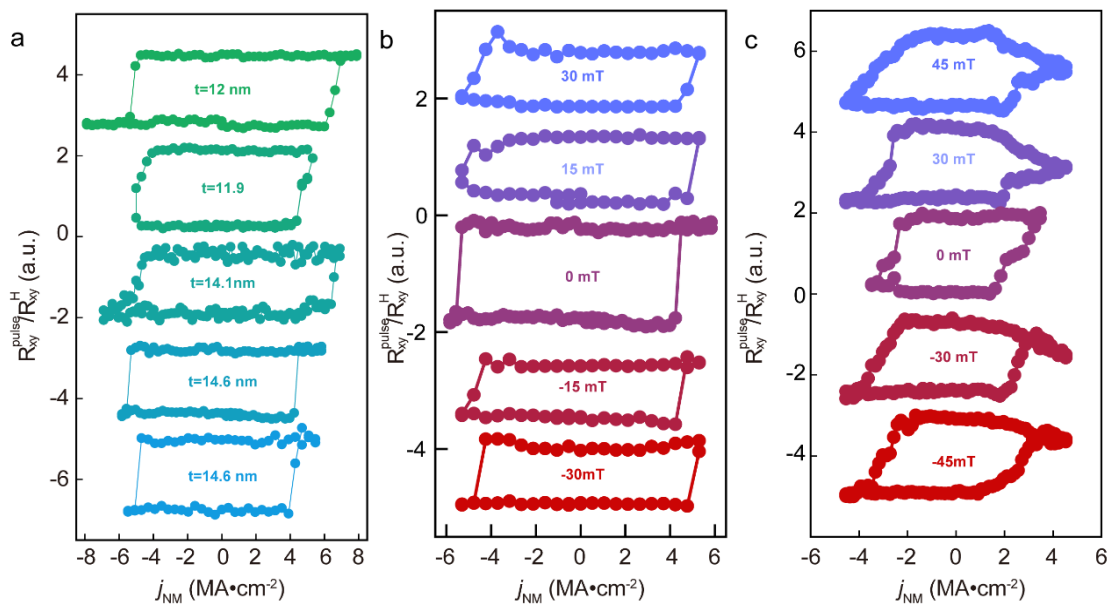

**Fig. S 14 | Repeatability of field-free SOT switching in NbIrTe$_4$ devices of varying thicknesses. a,** Representative field-free deterministic SOT switching results in devices with $t_{NbIrTe4} \approx 14$ nm. **b,c,** Repeatability of field dependent SOT switching behaviors on device with $t_{NbIrTe4}=12$nm (**b**) and device with $t_{NbIrTe4}=14.6$ nm (**c**). In both devices, the SOT switching polarity remains unaffected by an external $\mu_0 H_x \sim 45$ mT, indicating that the OOP spin torque is dominant in the magnetization switching process.

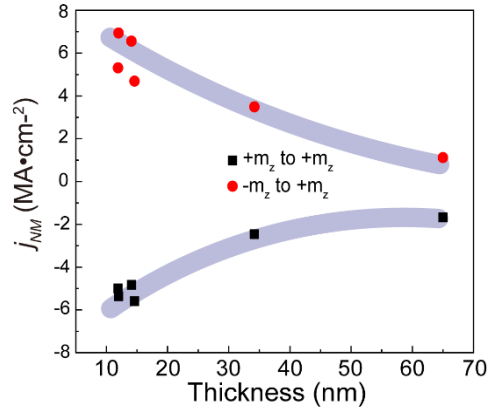

**Fig. S 15 | Threshold current for field-free SOT switching ($j_{NM}$) as a function of NbIrTe$_4$ thickness.**

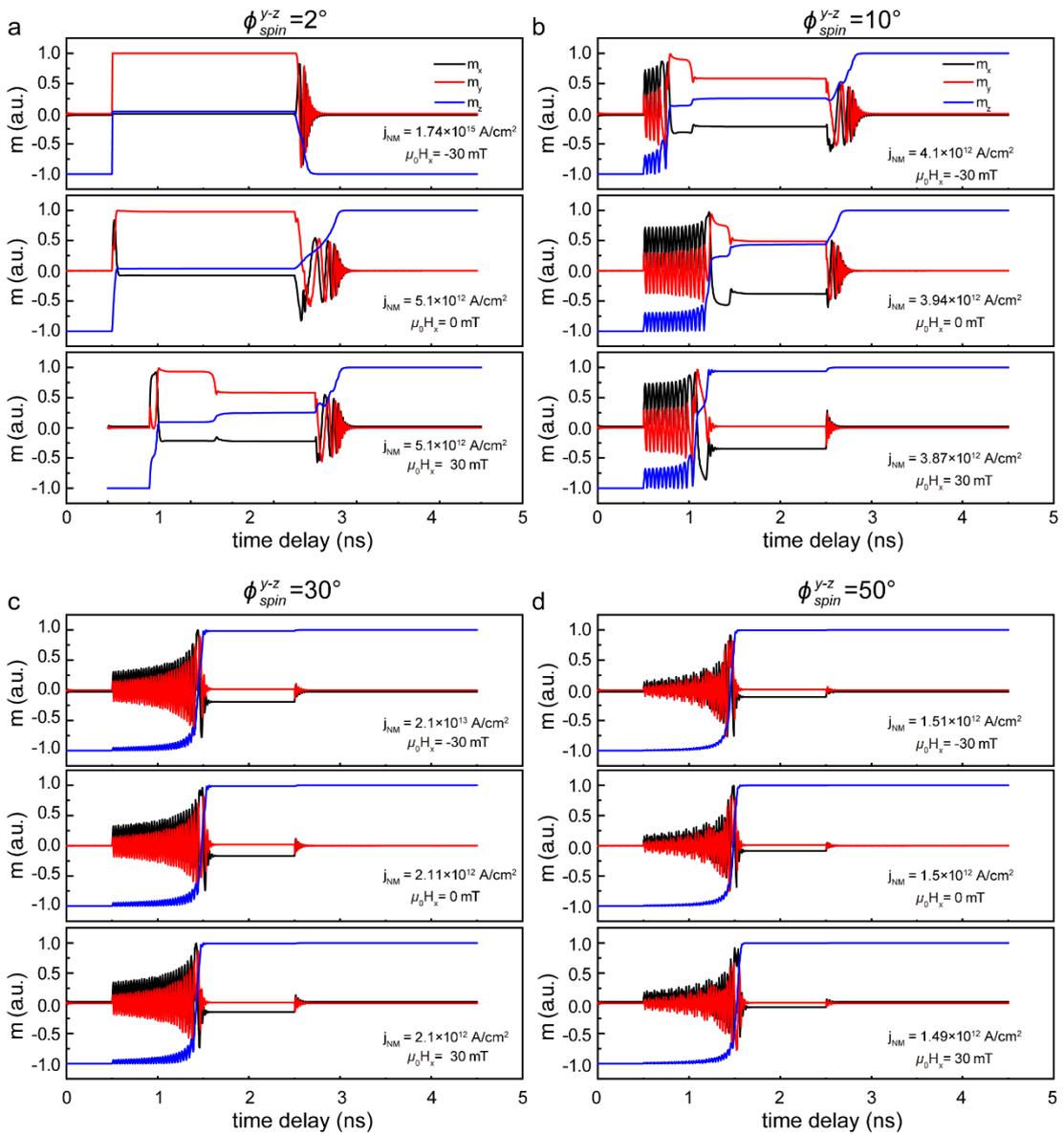

**Fig. S 16 | Spin dynamics under different spin title angle $\varphi_{spin}^{y-z}$. a,** $\varphi_{spin}^{y-z}$=2°. **b,**

($\varphi_{spin}^{y\text{-}z}$=10°). **c,** $\varphi_{spin}^{y\text{-}z}$=30°. **d,** $\varphi_{spin}^{y\text{-}z}$=50°. The top, middle and bottom panels in (a)-(d) show micromagnetic simulation results with spin dynamics under external magnetic fields of -30, 0 and 30 mT, respectively.

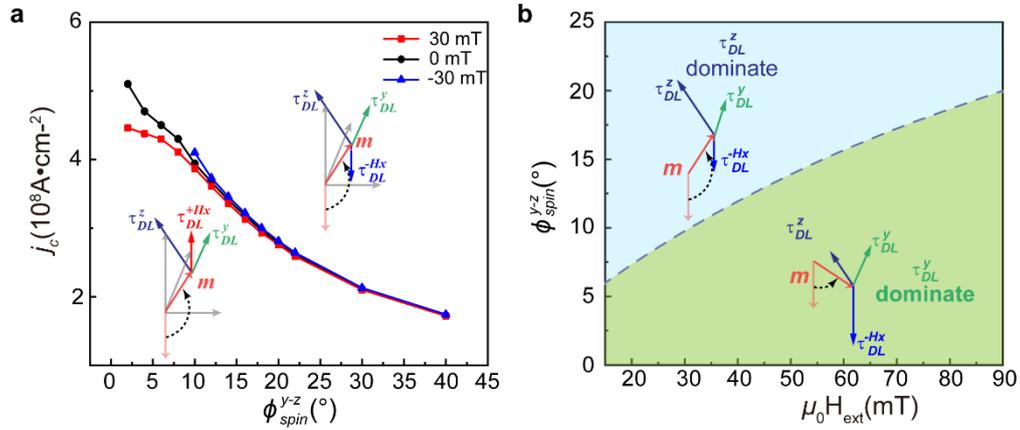

**Fig. S17 | Macrospin simulations showing the effect tilt angle of spin current ($\varphi_{spin}^{y-z}$) on magnetization switching dynamic a,** Threshold current densities required to switch the magnetization from $+m_z$ to $-m_z$ under various $\varphi_{spin}^{y-z}$ and in-plane magnetic fields ($\mu_0H_x$ = -30, 0, 30 mT). **b,** Phase diagram of $\varphi_{spin}^{y-z}$ and $\mu_0H_x$. In the blue (green) regions, $\tau_y^{DL}$ dominates and the SOT switching polarity depends on the in-plane field. In the blue region, $\tau_z^{DL}$ is dominant, making the SOT switching polarity independent of the in-plane field.

**Note1.  Raman Measurement of NbIrTe$_4$**

After depositing the ferromagnetic layer onto the two-dimensional substrate via magnetron sputtering, we performed Raman spectroscopy with minimum laser power (5%) to evaluate the structural integrity of the underlying material. Within the spectral range of 50–300 cm$^{-1}$, we identified nine prominent Raman peaks. This observation aligns closely with previously reported spectra for pristine two-dimensional NbIrTe$_4$[1], indicating that the sputtering process did not introduce significant lattice damage or alter the fundamental vibrational modes. The consistency of these Raman signatures with literature values also confirms that the structural quality and intrinsic symmetry of the NbIrTe$_4$ were preserved during the film growth, underscoring the robustness of our fabrication approach. Further angle-resolved Raman measurements (**Fig. 1c** in main body) revealed a pronounced anisotropic response to confirm the a-axis.

**Note2.  Comparison of ARPES measurement and DFT calculations**

To validate the electronic structure of bulk NbIrTe$_4$, we performed angle-resolved photoemission spectroscopy (ARPES) measurements and compared the results with density functional theory (DFT) calculations. **Fig. 2b** in the main text and **Fig. S2** in the supplementary information display the second-derivative ARPES spectra along the Γ- X and Γ- Y direction, respectively. Near the Fermi level (E$_F$), the ARPES spectra reveal a prominent 'blurred pocket', indicating significant bulk state contributions. Specifically, along the Γ–X direction, an elliptical electron pocket is observed near E$_F$, forming tilted energy cones. These features are consistent with the DFT-calculated band structure of bulk NbIrTe$_4$[4], as illustrated in **Fig. S3a**. At energies approximately 0.13 eV above E$_F$, our DFT results of bulk NbIrTe$_4$ further confirm the type-II nature of the band crossings between electron-like and hole-like pockets, leading to the formation of Weyl points. This correspondence between ARPES measurements and DFT calculations substantiates the three-dimensional Weyl semimetal topology of NbIrTe$_4$, which originates from its unconventional bulk electronic structure.

After determining the band structure, we built the wannier function based on the DFT results. **Fig. S3a** plots the the DFT and Wannier bands for comparison, which are consistent very well. Then, we used the Kubo formula to calculate the spin Hall conductance:

$$\sigma_{\alpha\beta}^{\text{SHC},\gamma} = \frac{-e\hbar}{N_k V_c} \sum_{\mathbf{k}} \sum_{n,m} (f_{n\mathbf{k}} - f_{m\mathbf{k}}) \frac{\text{Im}\left[\langle \psi_{n\mathbf{k}}|\frac{1}{2}\{s^\gamma, v_\alpha\}|\psi_{m\mathbf{k}}\rangle \langle \psi_{m\mathbf{k}}|v_\beta|\psi_{n\mathbf{k}}\rangle\right]}{(\varepsilon_{n\mathbf{k}} - \varepsilon_{m\mathbf{k}})^2 - (\hbar\omega + i\eta)^2} \quad (S1)$$

where $s^\gamma$ and $v_\alpha$ ($\alpha$, $\beta$, $\gamma = x, y, z$) are spin and velocity operators respectively. $\hat{j}_\gamma^\alpha = \frac{1}{2}\{s^\gamma, v_\alpha\}$ is the spin current operator for a spin current polarized along γ and transport along α direction, $f_{n\mathbf{k}}$ is the Fermi–Dirac distribution function for band $n$ and wave vector **k**. In the d.c. limit ($\omega = \eta = 0$). **Fig. S3b, c** plots the calculation result when current is applied along the a-axis and b-axis of bulk NbIrTe$_4$. the calculated $\sigma_{sh}$ along the a-axis of bulk NbIrTe$_4$ (**Fig. S3b**) includes components polarized along both the z- and y-directions, whereas along the b-axis only the y-polarized component appears (Figure S6). The calculated spin Hall conductance ($\sigma_{sh}$) of bulk NbIrTe$_4$ along the a-axis) reveals contributions polarized with both $\hat{\sigma}_y$ and $\hat{\sigma}_y$, while along the b-axis, only the component of $\hat{\sigma}_y$ is observed (**Fig. S3c**). This pattern matches the crystal symmetry, of with NbIrTe$_4$, where the mirror plane lies in the b-c plane.

**Note3. Band evolution of NbIrTe$_4$ from monolayer to bulk**

The bulk band structure of NbIrTe4, as shown in **Fig. S3a**, reveals features consistent with a Type-II Weyl semimetal, characterized by the tilted cones and Weyl point located near the Fermi level. In contrast, **Fig. S4a** illustrates the band structure of monolayer NbIrTe$_4$, which exhibits an energy gap compared to the bulk. The significant bandgap observed in the monolayer, driven by strong spin-orbit coupling, identifies monolayer NbIrTe$_4$ as a quantum spin Hall (QSH) insulator, analogous to its family TaIrTe$_4$[5]. Close to the Fermi level, the monolayer displays pronounced van Hove singularities (VHSs), further emphasizing the distinct electronic properties arising from reduced dimensionality and the enhanced quantum confinement effects. To further demonstrate this evolution, we calculated the band structures for NbIrTe$_4$ with layer numbers ranging from two to four layers, as depicted in **Figure S4b–d**. With increasing layer number, the band structure near the Γ point gradually develops elliptical pockets around E$_F$. This progression underscores the dimensional dependence of the topological features in NbIrTe$_4$ transitioning from a gapped QSH insulator in the monolayer to a Type-II Weyl semimetal in the bulk.

These band evolution characteristics have profound implications for the topological and spin transport properties of NbIrTe$_4$. To explore this, we constructed Wannier functions to calculate and compare the spin Berry curvature band projections for both monolayer and bulk NbIrTe$_4$, as shown in **Fig. S5**. The results reveal significant changes in the distribution and magnitude of the spin Berry curvature between the monolayer and bulk, reflecting the altered topological states.

Additionally, we calculated the spin Hall conductance (SHC) for NbIrTe$_4$ layers ranging from one to four layers (**Fig S6**). As illustrated in **Fig S7**, the out-of-plane spin Hall conductance ($\sigma_{sh}^z$) at the Fermi level increases with layer thickness, while the in-plane spin Hall conductance ($\sigma_{sh}^y$) decreases. Notably, in three and four-layer NbIrTe$_4$, the $\sigma_{sh}^z$ becomes dominant. This trend is in excellent agreement with our experimental observations using spin-torque ferromagnetic resonance (ST-FMR), where thinner NbIrTe$_4$ layers exhibit predominantly in-plane SHC, whereas thicker layers demonstrate dominant $\sigma_{sh}^z$ (see **Extended data Fig. 1**).

**Note4. Spin-torque ferromagnetic resonance measurement**

When the radiofrequency (RF) current ($I_{RF}$) is applied, the spin source layer generates an oscillating spin current that is injected into the adjacent CoFeB layer, which possesses magnetization with in-plane anisotropy. This spin current induces an oscillating spin-orbit torque (SOT) on the CoFeB layer, causing the magnetization direction and thus the bilayer resistance to oscillate due to the anisotropic magnetoresistance (AMR) of CoFeB. The mixing of the RF current and the time-varying resistance is detected as a voltage signal (V$_{mix}$) by a lock-in amplifier under an externally applied in-plane magnetic field $H_{ext}$ is applied. The voltage signal V$_{mix}$ from STFMR can be expressed as:

$$V_{mix} = SF_S(H) + AF_A(H) \qquad (S2)$$

where $F_S = \frac{\Delta H^2}{[\Delta H^2 + (H_{ext} - H_r)]}$ and $F_A = \frac{\Delta H(H_{ext} - H_r)}{[\Delta H^2 + (H_{ext} - H_r)]}$ are the symmetric and antisymmetric Lorentzian functions, respectively. S is the amplitudes of the symmetric $F_S$ signals and is proportional to the current-induced in-plane torque $\tau_\parallel$:

$$V_s = -\frac{1}{4}\frac{dR}{d\theta}\frac{\gamma I_{RF}\cos\theta}{\Delta H}\tau_\parallel \qquad (S3)$$

where θ is the angle between $H_{ext}$ and the injection direction of radiofrequency current $I_{RF}$. $\tau_\parallel$ contains in-plane spin current $\sigma_y$ induced damping-like torque

$\tau_{DL,y} (\propto m \times \sigma_y \times m)$ and out-of-plane spin current $\sigma_z$-induced field like torque $\tau_{FL,z} (\propto m \times \sigma_z)$. On the other hand, A is the amplitudes the antisymmetric $F_A$ signals and is proportional to the current-induced out-of-plane torque $\tau_\perp$:

$$V_A = -\frac{1}{4}\frac{dR}{d\theta}\frac{\gamma I_{RF}\cos\theta}{\Delta H}\tau_\perp \qquad (S4)$$

where $\tau_\perp$ contains Oersted-field torque $\tau_{Oe}$ induced by $I_{RF}$, the field-like torque $\tau_{FL,y}(\propto m \times \sigma_y)$ induced by the in-plane spin current $\sigma_y$ and the amping-like torque $\tau_{DL,z}(\propto m \times \sigma_z \times m)$ induced by out-of-plane spin current $\sigma_z$. By fitting the angle dependence of the $V_S$ and $V_A$:

$$\begin{aligned}S &= S_{DL}^X \sin\phi\sin 2\phi + S_{DL}^Y \cos\phi\sin 2\phi + S_{FL}^Z \sin 2\phi \\ A &= A_{FL}^X \sin\phi\sin 2\phi + A_{FL}^Y \cos\phi\sin 2\phi + A_{DL}^Z \sin 2\phi\end{aligned} \qquad (S5)$$

where $S_{DL}^X$, $S_{DL}^Y$, and $A_{DL}^Z$ is the coefficients for the damping-like torque for x, y, z polarized spin current; $A_{FL}^X$ $A_{FL}^Y$ and $S_{FL}^Z$ are the corresponding field-like torques for x, y, z polarized spin current respectively. By consider the $A_{FL}^Y$ due to the Oersted field as $A_{FL}^Y = H_{rf}[1 + 4\pi M_{eff}/H_{ext}]^{1/2}$, we can estimate the effective charge-to-spin conversion efficiency $\theta_{sh}$ of $\sigma_y$ and $\sigma_z$ can be as:

$$\begin{aligned}\theta_{sh} &= \frac{S_{DL}^Y}{A_{FL}^Y}\frac{e\mu_0 M_s t d}{\hbar}\sqrt{1+(4\pi\mu_0 M_{eff}/H_r)} \\ \theta_{sh,y} &= \frac{S_{DL}^X}{A_{FL}^Y}\frac{e\mu_0 M_s t d}{\hbar}\sqrt{1+(4\pi\mu_0 M_{eff}/H_r)} \\ \theta_{sh,z} &= \frac{A_{DL}^Z}{A_{FL}^Y}\frac{e\mu_0 M_s t d}{\hbar}\sqrt{1+(4\pi\mu_0 M_{eff}/H_r)}\end{aligned} \qquad (S6)$$

where t and d are the thickness of the ferromagnetic and spin source layer, respectively. The effective magnetization $\mu_0 M_{eff}$ is determined by fitting $f$ vs. $H_r$ to the Kittel equation:

$$f = \left(\frac{\gamma}{2\pi}\right)\mu_0\sqrt{(H_r - H_k)(H_r - H_k + M_{eff})} \qquad (S7)$$

The effective Gilbert damping constant α is obtained by a linear fit of $\Delta H$ vs. $f$ using

$$\Delta H = \Delta H_0 + \frac{2\pi\alpha f}{\gamma} \qquad (S8)$$

Taking the charge conductance estimated from parallel model (Eq. S2), the spin Hall conductivity of $\sigma_y$ and $\sigma_z$ can be obtained according to:

$$\sigma_{sh,y(z)} = \sigma_c \theta_{sh,y(z)}. \qquad (S9)$$

We performed angle dependent STFMR measurement on devices with $t_{NbIrTe4}$ varing from 12 to 119.6nm. **Fig. S9** shows the representative STFMR measurement signal in device with $t_{NbIrTe4}$=119.2 nm, where the noticeable change in the $V_A$ component between $+\mu_0 H_{ext}$ and $-\mu_0 H_{ext}$ indicates the OOP spin current contribution. The effective magnetization $\mu_0 M_{eff}$ was determined as 0.7 $\pm$ 0.01 T and the Gilbert damping constant $\alpha$ of CoFeB was determined as 0.0056 $\pm$0.0004. **Fig. S10** shows the angle dependent antisymmetric ($V_A$) and symmetric ($V_S$) components for devices with $t_{NbIrTe4}$ ranging from 14 to 119.6 nm. By fitting the them to Eq. S7, we can get the thickness dependent spin hall conductance, as shown in **Fig. 3c** and **Table S1**.

### Note5. Anomalous Hall effect loop shift measurements

To evaluate the out-of-plane (OOP) spin current $\sigma_z$ and its corresponding damping-like torque $\tau_{DL,z}$ in our NbIrTe4-based devices, we examined the anomalous Hall effect (AHE) loop shifts at different current densities. As shown in **Fig. S11**, we initially performed AHE measurements at a low DC current of 0.1 mA to confirm the strong perpendicular magnetic anisotropy (PMA). The square loop obtained by sweeping $H_z$ reveals robust PMA, and an anisotropy field of $H_a \approx$ 250 mT is identified from the from the $H_x$ sweep.

In NbIrTe4, the mirror symmetry does not exist in the ac plane. When current is applied along the a-axis, the $\sigma_z$ and an associated damping-like torque $\tau_{DL,z}$ are allowed. To directly probe $\sigma_z$ and $\tau_{DL,z}$ in our NbIrTe4/Ti/CoFeB samples with PMA, we used current pulse (100 μs pulse width) of different amplitudes to perform the AHE measurement. As displayed in **Fig. S1212**, when the positive/negative currents applied along the a-axis exceed the threshold value of 8 mA (~$j_c$= 1.2 MA/cm$^2$ for device with $t_{NbIrTe4}$= 34.2 nm), the switching point of AHE loop starts to shift to the left/right, respectively, which indicates that the down-spin is generated[2, 3].

The current-induced out-of-plane effective field $\Delta\mu_0 H_{z,s}$ can be extracted from the shift of AHE loop measurements. The equation can be expressed as:

$$\Delta\mu_0 H_{z,s} = \frac{|H_{\text{shift}}(I^+) - H_{\text{shift}}(I^-)|}{2} \tag{S10}$$

where $H_{\text{shift}}(I^+)$ and $H_{\text{shift}}(I^-)$ are the shift of AHE loop at positive and negative current respectively. The extracted $\Delta\mu_0 H_{z,s}$ is plot in **¡Error! No se encuentra el origen de la referencia.13**, which increases linearly. For the measured device, we have all structure as NbIrTe$_4$(34.2nm) /Ti(2nm) /CoFeB(1.3nm) /MgO(2nm)/Ta(1nm), the conductivities of 34.2 nm NbIrTe$_4$ and the PMA structure, Ti (2 nm)/CoFeB (1.3 nm)/MgO (2 nm)/Ta (1nm) layer, can be obtained from standard four-probe measurements as $\rho_{\text{NbIrTe4}}$ =226.93 μΩ·cm and $\rho_{\text{PMA}}$ = 134.9 μΩ·cm. The $j_{\text{NM}}$ can be calculated with the parallel current model:

$$j_{NM} = I \frac{\frac{\rho_{PMA}}{t_{PMA}}}{\frac{\rho_{PMA}}{t_{PMA}} + \frac{\rho_{NM}}{t_{NM}}} \times \frac{1}{w \times t_{NM}} \tag{S11}$$

By fitting the linear relationship of $\Delta\mu_0 H_{z,s}$ and $j_c$, we get the efficiency of current induced OOP field as $\Delta\mu_0 H_{z,s}/j_{NM}$ ~0.29±0.01 mT/ (10$^{10}$A/m). Then the charge to spin conversion efficiency of σ$_z$ can be estimated as $0.06 \pm 0.002$ from follow equation:

$$\theta_{DL} = \frac{J_{SH}}{J_{NM}} = \frac{2e\mu_0 M_s t_{FM}}{\hbar} \frac{\Delta H_{z,s}}{\Delta j_c} \tag{S12}$$

where $M_s$=0.533×10$^6$ A/m is the saturated magnetism measured by vibrating magnetometer and $t_{FM}$ is the thickness of the CoFeB laye. The extracted value of $\theta_{sh}^z$ aligns well with the results from our ST-FMR measurements, reinforcing the conclusion that NbIrTe$_4$ effectively generates an out-of-plane spin current.

### Note6. Repeatability of Field-Free SOT Switching

To confirm the reproducibility and robustness of the deterministic switching behavior driven by out-of-plane ($\hat{\sigma}_z$) spin current, we performed switching measurements on multiple NbIrTe$_4$-based devices of varying thicknesses. As shown in

**Fig. S 1414**, all devices exhibit field-free SOT switching, confirming the reproducibility of the out-of-plane torque mechanism across different thicknesses. It is noteworthy that for devices with $t_{NbIrTe4}$ around 14 nm, the switching current density remains consistently around ~4.5 MA/cm², indicating robust and stable out-of-plane spin current-induced switching in this thickness range. Here, the current density $j_{NM}$ is determined with the parallel current model using Eq. S2.

**Fig. S14b, c** demonstrates the repeatability of the SOT switching behavior across devices with different NbIrTe$_4$ thicknesses. In both devices, the SOT switching polarity remains unaffected by an external in-plane magnetic field of approximately 45 mT. This consistent switching polarity, irrespective of the external field direction, underscores the dominance and robustness of the out-of-plane spin torque in the magnetization switching process. Such field insensitivity confirms that the out-of-plane spin currents are the primary driving force behind the deterministic, field-free switching observed in our NbIrTe$_4$/Ti/CoFeB devices.

**Note7. Thickness dependent critical current for field free SOT switching**

The efficiency of OOP spin current generation in NbIrTe$_4$ is significantly influenced by the material's thickness. As the NbIrTe$_4$ layer becomes thicker, the generation efficiency of OOP spin currents increases, leading to a reduction in the threshold current density required for SOT-induced switching. This relationship is systematically explored through a combination of experimental measurements and theoretical simulations, as depicted in **Fig. S15** and **Fig. S17**.

Furthermore, by integrating the findings from **Fig 4a** with results from STFMR measurement (**Fig 3c**), we establish a clear relationship between the spin tilt angle ($\varphi_{spin}^{y-z}$) and the threshold current density ($j_{NM}$) across different NbIrTe$_4$ thicknesses (**Fig 4b**). The spin tilt angle $\varphi_{spin}^{y-z}$ was derived from angle-dependent ST-FMR measurements, while $j_{NM}$ was obtained from SOT switching measurement on Hall devices.

Combining the insights from macrospin simulation (**Fig S17a**) with experimental data presented in **Figure S15**, we observe a consistent trend that the increasing the spin tilt angle ($\varphi_{spin}^{y-z}$) correlates with a reduction in the threshold current density ($j_{NM}$). The experimental data (red line) and micromagnetic simulation results (black line) both demonstrate that as the NbIrTe$_4$ thickness increases, thereby enhancing the spin tilt angle, the efficiency of OOP spin current generation improves. This enhancement results in lower $j_{NM}$, confirming that thicker NbIrTe$_4$ layers facilitate more effective OOP spin torque, thereby enabling more efficient and reliable magnetization switching.

### Note8. Spin dynamics under OOP spin current

To elucidate the influence of spin tilt angle $\varphi_{spin}^{y-z}$ on magnetization dynamics driven by out-of-plane (OOP) spin currents, we performed micromagnetic simulations with $\varphi_{spin}^{y-z}$ ranging from 2° to 50°, as depicted in **Figure S16**.

At small spin tilt angles ($\varphi_{spin}^{y-z}$ = 2°, **Figure S 16a**), the in-plane SOT dominates the magnetization dynamics. The magnetization aligns predominantly along the y-axis under the influence of the in-plane component of the spin current. Subsequently, the external magnetic field determines the switching polarity of the perpendicular magnetization, that is, the switching from -m$_z$ to +m$_z$ cannot happened under positive field (only for zero/negative field). Consequently, the switching process is significantly influenced by the external magnetic field direction, making it susceptible to external perturbations and limiting the robustness of deterministic, field-free switching. As the $\varphi_{spin}^{y-z}$ increases, the OOP-SOT becomes increasingly dominant. As shown in **Figure S 16b-d**, the magnetization undergoes a direct switching to the z-axis, effectively bypassing the reliance on external magnetic fields for determining switching polarity. The inherent perpendicular magnetic anisotropy (PMA) of the CoFeB layer stabilizes the magnetization in the z-direction, completing the magnetization reversal process. Enhanced OOP-SOT leads to reduced magnetic damping and a higher proportion of magnetization flips driven by OOP spin currents, resulting in faster switching speeds and lower threshold current densities. This transition facilitates the realization of high-density, high-speed, and low-power spintronic memory devices, as the switching becomes more efficient and less dependent on external field conditions.

**Fig. S17a** presents the threshold current density required to switch the magnetization from **+m$_z$** to **-m$_z$** driven by SOT for various tilt angles and external in-plane magnetic fields (μ$_0$H$_x$=-30, 0 and 30mT). For large tilt angles, the $\hat{\sigma}_z$ exerts a dominant torque $\tau_z^{DL}$ ($\propto \mathbf{m} \times \hat{\sigma}_z \times \mathbf{m}$), allowing direct one-step switching of **+m$_z$** to -**m$_z$** regardless of $\pm H_x$. Conversely, in the case of smaller $\varphi_{spin}^{y-z}$, the $\tau_y^{DL}$ ($\propto \mathbf{m} \times \hat{\sigma}_y \times \mathbf{m}$) from $\hat{\sigma}_y$ dominates, causing **+m$_z$** to align along the *y*-direction before switching. In this case, the magnetization switching depends on the relative orientation of $\tau_{\pm Hx}$ and $\tau_z^{DL}$. In our simulation, $\tau_{+(-)Hx}$ aligns (anti-)parallel to $\tau_z^{DL}$, the reversal is assisted (hindered), and the corresponding threshold current density for switching decreases (increases). Conversely, in the case of smaller $\varphi_{spin}^{y-z}$, the interaction of $\tau_y^{DL}$ and $\tau_{+(-)Hx}$ dominate $\tau_{+(-)Hx}$ and the switching direction is determined by direction of $H_x$, that is, no switching occurs under -$H_x$ applied.

By numerically simulating switching under different magnetic fields, we obtain the phase diagram of spin tilt angle $\varphi_{spin}^{y-z}$ and $H_x$ in F**ig S17b**, where the green (red) region the $\tau_z^{DL}$ ($\tau_y^{DL}$) is dominant, and the polarity of the magnetization reversal is affected by the in-plane magnetic field; in the blue region, the $\tau_z^{DL}$ is dominant, and the polarity of the magnetization reversal is not affected by the in-plane magnetic field. According to our STFMR results, the spin tilt angle $\varphi_{spin}^{y-z}$ of the device with $t_{NbIrTe4}$=37.6nm is 35°, so it could be understood that the switching polarity will not change within range of $H_x \sim \pm 60$mT.

### Note9. Estimation of power consumption for SOT switching

The power consumption of a device during spin-orbit torque (SOT) switching can be estimated using the following relationship:

$$p = \rho J_c^2 \qquad (S13)$$

where $p$ is the power consumption, $\rho$ is the resistivity of the current channel, $J_c$ is the current density required to induce magnetization switching. The Oersted field $H_O$ generated by the current can be approximated using Ampère's Law for a long wire:

$$\mu_0 H_O = \mu_0 \frac{I_c}{2r} \cong \frac{\mu_0 j d}{2} \qquad (S14)$$

where $I_c$ is the current through the wire and r is the distance from the center of the current-carrying wire to the point where the magnetic field is measured. For an approximation near the wire, $r$ would typically be the width (W>>d) of the conductive channel. Taking the resistivity of the Cu channel, and thickness of 10nm, the power consumption due to the current flowing through the material can be calculated using:

$$p = j^2 \rho = \left(\frac{2H_O}{d}\right)^2 \rho \qquad (S15)$$

**Supplemental Table**

We summarize the charge-to-spin conversion efficiency of z-polarized spin currents recently observed in low-symmetry materials (as shown in Table S1). The research systems in these studies are categorized into two groups: layered materials with broken lattice symmetry (space group *Pmn2₁*), and non-collinear antiferromagnetic materials, where mirror symmetry is broken by the magnetic structure. Our results demonstrate the highest spin Hall conductance for z-polarized spins, with a significantly larger ratio of z-spin to y-spin conversion compared to other materials.

**Table S1. Summary of charge to spin conversion efficiency**

| Materials | Type | $\sigma_{sh,y}$ ($10^5$ $\hbar/2e\cdot\Omega^{-1}m^{-1}$) | $\sigma_{sh,z}$ ($10^5$ $\hbar/2e\cdot\Omega^{-1}m^{-1}$) | $\sigma_{sh,z}/\sigma_{sh,y}$ | Thickness of NM $t_{NM}$ | **Ref** |
|---|---|---|---|---|---|---|
| **TaIrTe₄** | Layer Material (*Pmn2₁*) | 0.539 | 0.205 | 0.113 | 25 | [6] |
| | | 0.246 | 0.147 | 0.6 | 11 | [7] |
| **WTe₂** | | 0.079 | 0.034 | 0.45 | 5.5 | [8] |
| | | 0.074 | 0.018 | 0.24 | 5 | [9] |
| | | 0.35 | 0.021 | 0.06 | 2.6 | [10] |
| | | 0.13 | 0.03 | 0.23 | 8 | [11] |
| **β-MoTe2** | | 0.058 | 0.01 | 0.18 | 0.7 | [12] |
| **NbIrTe₄** | | 1.75±0.13 | 0.35±0.08 | 0.20 | 14 | **Our Work** |
| | | 1.57±0.10 | 0.26±0.09 | 0.17 | 17.7 | |
| | | 1.45±0.09 | 0.33±0.01 | 0.23 | 25.3 | |
| | | 0.83±0.05 | 0.26±0.07 | 0.31 | 37.6 | |
| | | 1.00±0.08 | 0.26±0.03 | 0.26 | 39.8 | |
| | | 0.69±0.06 | 0.22±0.05 | 0.32 | 40.8 | |
| | | 0.15±0.09 | 0.48±0.02 | 3.29 | 53.8 | |
| | | 0.12±0.03 | 0.43±0.04 | 3.50 | 51.3 | |
| | | 0.31±0.09 | 1.25±0.11 | 4.05 | 63.8 | |
| | | 0.26±0.06 | 1.17±0.03 | 4.56 | 92.1 | |
| | | 0.25±0.09 | 1.17±0.13 | 4.62 | 119.2 | |
| **MnGaN** | Anti-Ferromagnetic | 0.11 | 0.08 | 0.72 | 20 | [13] |
| **MnPd₃** | | 5.13 | 0.11 | 0.026 | 15 | [14] |
| **RuO₂** | | 0.178 | 0.07 | 0.39 | 6 | [15] |

# References:

1. Zhang, J. et al. Colossal Room-Temperature Terahertz Topological Response in Type-II Weyl Semimetal NbIrTe$_4$. *Adv. Mater.* **34**, 2204621 (2022).

2. Kao, I. et al. Deterministic Switching of a Perpendicularly Polarized Magnet Using Unconventional Spin–Orbit Torques in WTe$_2$. *Nat. Mater.* **9**, 1029-1034 (2022).

3. Baek, S. C. et al. Spin Currents and Spin–Orbit Torques in Ferromagnetic Trilayers. *Nat. Mater.* **17**, 509-513 (2018).

4. Lee, J. et al. Spin-Orbit-Splitting-Driven Nonlinear Hall Effect in NbIrTe$_4$. *Nat. Commun.* **15**, 3971 (2024).

5. Tang, J. et al. Dual Quantum Spin Hall Insulator by Density-Tuned Correlations in TaIrTe$_4$. *Nature* **628**, 515-521 (2024).

6. Liu, Y. et al. Field-Free Switching of Perpendicular Magnetization at Room Temperature Using Out-of-Plane Spins from TaIrTe$_4$. *Nat. Electron.* **6**, 732-738 (2023).

7. Zhang, Y. et al. Room Temperature Field-Free Switching of Perpendicular Magnetization through Spin-Orbit Torque Originating from Low-Symmetry Type II Weyl Semimetal. *Sci. Adv.* **9**, eadg9819 (2023).

8. MacNeill, D. et al. Control of Spin–Orbit Torques through Crystal Symmetry in WTe$_2$/Ferromagnet Bilayers. *Nat. Phys.* **13**, 300-305 (2016).

9. Shi, S. et al. Observation of the Out-of-Plane Polarized Spin Current from CVD Grown WTe$_2$. *Adv. Quantum Technol.* **4**, 2100038 (2021).

10. Wang, X. et al. Room Temperature Field-Free Switching of CoFeB/MgO Heterostructure Based on Large-Scale Few-Layer WTe$_2$. *Cell Rep. Phys. Sci.* **4**, 101468 (2023).

11. Wang, F. et al. Field-Free Switching of Perpendicular Magnetization by Two-Dimensional PtTe$_2$/WTe$_2$ Van Der Waals Heterostructures with High Spin Hall Conductivity. *Nat. Mater.* (2024).

12. Li, R. et al. Layer-Dependent Spin-Orbit Torques Generated by the

Centrosymmetric Transition Metal Dichalcogenide β−MoTe$_2$. *Phys. Rev. B* **100**, 184402 (2019).

13. Nan, T. et al. Controlling Spin Current Polarization through Non-Collinear Antiferromagnetism. *Nat. Commun.* **11**, 4671 (2020).

14. DC, M. et al. Observation of Anti-Damping Spin–Orbit Torques Generated by in-Plane and Out-of-Plane Spin Polarizations in MnPd3. *Nat. Mater.* **22**, 591-598 (2023).

15. Bose, A. et al. Tilted Spin Current Generated by the Collinear Antiferromagnet Ruthenium Dioxide. *Nat. Electron.* **5**, 267-274 (2022).